\documentclass[journal, twocolumn]{IEEEtran}
\usepackage{tikz,tikz-qtree}	
\usepackage{graphicx,cite}
\usepackage{amsmath,amssymb,bbm}
\usepackage{algorithm,algpseudocode}
\usepackage{rotating,multirow}

\newcommand{\bld}[1]{{\bf #1}}
\newcommand{\bs}[1]{\boldsymbol{#1}}
\newcommand{\Expectation}[1]{\mathbb{E} \left\lbrace #1 \right\rbrace }

\newcommand{\inCurly}[1]{ \left\lbrace #1 \right\rbrace}
\newcommand{\inBrackets}[1]{ \left( #1 \right)}
\newcommand{\inSqBrackets}[1]{ \left[ #1 \right]}
\newcommand{\given}{ \left|\right.}

\newcommand{\boundellipse}[4]
{(#1) ellipse [x radius = #2, y radius = #3, rotate = #4]}

\newcommand{\openRectangle}[5]
{ 				(#1,#2) 
                -- (#1,#2-#4) 
                -- (#1+#3,#2-#4)node[below,midway] {#5}
                -- (#1+#3,#2)}
\newcommand{\openRectangleDown}[5]
{ 				(#1,#2) 
                -- (#1,#2+#4) 
                -- (#1+#3,#2+#4)node[above,midway] {#5}
                -- (#1+#3,#2)}
                
\newcommand{\vArray}[3]
{ \matrix(m)[matrix of nodes,nodes in empty cells,
  nodes={inner sep=1pt,draw={#1},line width=0.2pt,text width=10pt, text height=10pt,text depth=0.1pt},
  row sep=-0.5\pgflinewidth,
  column sep=-\pgflinewidth,
  ]at (#2,#3){
    $\mu_1$\\
    $\mu_2$\\
     \\
     \\
     \\
     \\
     \\
     \\
   $\mu_K$\\
  };				}   

\newcommand{\mic}[3]
{
		\begin{scope}[shift={#1},rotate=#2,scale=#3]		
			\draw (0,0) -- (0:1);
			\draw (1.2,0) circle (.2);
			\draw (1,-0.25) -- (1,0.25); 			
		\end{scope}
}

\newcommand{\speakerP}[3]
{
	\begin{scope}[shift={#1},scale=#2]
		\draw[#3] (-0.7,1.2) circle (0.45);
		\draw[thick,#3] (0,0) arc(0:180:0.7);
	\end{scope}
}

\newcommand{\speakerF}[3]
{
	\begin{scope}[shift={#1},scale=#2,rotate=#3]
			\draw[] (0,0) rectangle (0.75,1);
			\draw[line width=0.05mm] (0.75/2,0.85) circle(0.1cm);
			\draw[line width=0.05mm] (0.75/2,0.85) circle(0.06cm);
			
			\draw[line width=0.05mm] (0.75/2,0.35) circle(0.1cm);			
			\draw[line width=0.05mm] (0.75/2,0.35) circle(0.22cm);
			\draw[line width=0.2mm] (0.75/2,0.35) circle(0.3cm);
			
			\draw[line width=0.05mm] (0.05,0.05) circle(0.02cm);
			\draw[line width=0.05mm] (0.7,0.95) circle(0.02cm);
			\draw[line width=0.05mm] (0.05,0.95) circle(0.02cm);
			\draw[line width=0.05mm] (0.7,0.05) circle(0.02cm);			
									
	\end{scope}
}

\newcommand{\speakerM}[3]
{
	\begin{scope}[shift={#1},scale=#2,rotate=#3]
			\draw[fill=gray] (0.5,0.05) ellipse [x radius = 0.08, y radius = 0.15];
			\draw[fill=gray] (-0.5,0.05) ellipse [x radius = 0.08, y radius = 0.15];
			
			\draw[fill=gray] (0.0,0.6) ellipse [x radius = 0.05, y radius = 0.1];
			
			\draw[fill=gray] (0,0) ellipse [x radius = 0.5, y radius = 0.6];
			\draw[line width=0.05mm,fill=gray] (-0.15,0.3) ellipse [x radius = 0.05, y radius = 0.1];
			\draw[line width=0.05mm,fill=gray] (0.15,0.3) ellipse [x radius = 0.05, y radius = 0.1];
	\end{scope}
}

\title{Directional MCLP Analysis and Reconstruction for Spatial Speech Communication}
\author{Srikanth~Raj~Chetupalli, and Thippur V. Sreenivas,~\IEEEmembership{Senior member,~IEEE}\\
        Dept. of Electrical Communication Engineering, Indian Institute of Science, Bengaluru, 560012.}
\begin{document}

\maketitle

\begin{abstract}
 Spatial speech communication, i.e., the reconstruction of spoken signal along with the relative speaker position in the enclosure (reverberation information) is considered in this paper. Directional, diffuse components and the source position information are estimated at the transmitter, and perceptually effective reproduction is considered at the receiver. We consider spatially distributed microphone arrays for signal acquisition, and node specific signal estimation, along with its direction of arrival (DoA) estimation. Short-time Fourier transform (STFT) domain multi-channel linear prediction (MCLP) approach is used to model the diffuse component and relative acoustic transfer function is used to model the direct signal component. Distortion-less array response constraint and the time-varying complex Gaussian source model are used in the joint estimation of source DoA and the constituent signal components, separately at each node. The intersection between DoA directions at each node is used to compute the source position. Signal components computed at the node nearest to the estimated source position are taken as the signals for transmission. 

At the receiver, a four channel loud speaker (LS) setup is used for spatial reproduction, in which the source spatial image is reproduced relative to a chosen virtual listener position in the transmitter enclosure. Vector base amplitude panning (VBAP) method is used for direct component reproduction using the LS setup and the diffuse component is reproduced equally from all the loud speakers after decorrelation. This scheme of spatial speech communication is shown to be effective and more natural for hands-free telecommunication, through either loudspeaker listening or binaural headphone listening with head related transfer function (HRTF) based presentation.

\end{abstract}

\section{Introduction}
	\label{sec:intro}
	\par Communication is an integral part of human interaction and growth. The invention of telephone has enabled long-distance telecommunication of speech. With the advent of digital high bit-rate communication channels, the visual aspect of communication has also been integrated into telecommunication and also extended to multi-party conversations and distant conferencing. However, these types of speech communications are not yet as effective and natural feeling similar to a direct acoustic communication mainly because the signals do not carry the ``spatial dimension''. In traditional speech telecommunication, the source spectral attributes such as the speaker's identity and emotion are well preserved (depending on the channel bit-rate). But, the spatial attributes such as the source position, or its movement, and other properties of speaking enclosure and background, such as indoor/outdoor, small/big enclosures, noisy/quiet ambiance, etc. are not emphasized. 
\par ``Spatial speech communication'' (SSC) is the communication of natural quality speech signal along with its spatial attributes at the transmitter and its reasonably good spatial reproduction at the receiver for a chosen virtual listener (VL) position in the transmitter space. The perceived spatial attributes include the source position, reverberation, size, and other acoustic characteristics of the transmitter enclosure. Signal acquisition at the transmitter should be hands-free, and the source is free to move inside the enclosure. The listener is also free to choose a VL position in the transmitter enclosure and the transmitter acoustic scene should be reproduced for the selected VL position at the receiver. 
\begin{figure}[t]
	\begin{minipage}{0.48\linewidth}
		\centering
		\begin{tikzpicture}[scale=0.45,rotate=0,every node/.style={rotate=0}]				
		\draw[] (0,0) rectangle (8,8);				
		\speakerP{(3.25,4.5)}{0.4}{red};
		\mic{(2,0)}{90}{0.4};
		\mic{(2.4,0)}{90}{0.4};
		\draw[rounded corners=0.5mm] (1.8,0.3) rectangle (2.6,0.65);
		\mic{(6,8)}{-90}{0.4};
		\mic{(6.4,8)}{-90}{0.4};
		\draw[rounded corners=0.5mm] (5.8,7.35) rectangle (6.6,7.7);			
		\mic{(0,5.0)}{0}{0.4};
		\mic{(0,5.4)}{0}{0.4};
		\draw[rounded corners=0.5mm] (0.3,4.8) rectangle (0.65,5.6);			
		\mic{(8,2)}{180}{0.4};
		\mic{(8,2.4)}{180}{0.4};					
		\draw[rounded corners=0.5mm] (7.35,1.8) rectangle (7.7,2.6);			
		\speakerM{(3.75,2)}{0.8}{0};
		\end{tikzpicture}
		\centerline{\scriptsize Tx: Transmitter}
	\end{minipage}
	\begin{minipage}{0.48\linewidth}
		\centering
		\begin{tikzpicture}[scale=0.45,rotate=90,every node/.style={rotate=90}]
		\draw[] (0,0) rectangle (8,8);
		\speakerP{(3.25,5.5)}{0.4}{red,very thin,rotate=-90};		
		\speakerM{(3.75,3)}{0.8}{0};
		\speakerF{(0.15,0.9)}{1}{-90};
		\speakerF{(6.9,0.9)}{1}{-90};
		\speakerF{(6.9,7.9)}{1}{-90};
		\speakerF{(0.15,7.9)}{1}{-90};
		\end{tikzpicture}
		\centerline{\scriptsize Rx: Receiver}
	\end{minipage}
	\caption{Transmitter and receiver setup for spatial speech communication; (i) distributed node microphone arrays at Tx; (ii) independent signal and DoA estimation at each node; (iii) virtual listener selection by Rx.} \label{fig:txrxSetup}
	\vspace{-15pt}
\end{figure}
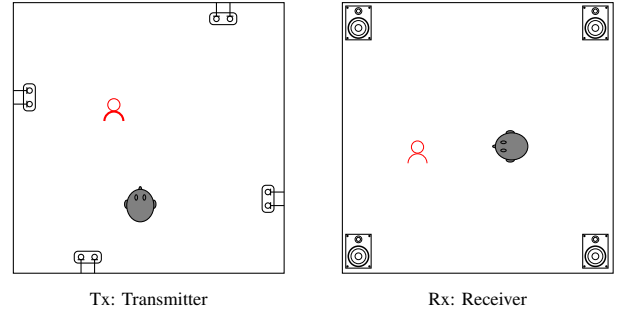
\par We consider a parametric spatial audio approach to spatial speech communication in the present work, aiming at perceptually transparent reconstruction. Inter-aural time, level differences (ITD, ILD), and inter-aural coherence (IC) cues in narrow frequency bands are identified as the important perceptual cues for the faithful reconstruction of a spatial audio scene\cite{Baumgarte2003Binaural, Parametric2015Kowalczyk, pulkki2007spatial}. Towards this, a parametric representation of the reverberant signal, which characterizes the source direction, and its directional and diffuse components, is found to be sufficient. Parametric approaches also allow for object-based coding of audio for efficient transmission, similar to the spatial audio object coding scheme \cite{breebaart2008spatial}, where the recorded acoustic scene is represented in terms of the source audio objects and their spatial attributes. A mono down-mix of the signal is transmitted along with the spatial attributes. Parametric approaches also allow for arbitrary loudspeaker arrangement for spatial reproduction (or even binaural reproduction for headset listening), flexible microphone placement for signal acquisition, and efficient coding for transmission and storage. Further, the receiver has the flexibility to modify the acoustic scene, i.e., to selectively suppress/enhance a specific source and change its position relative to the listener.
\par Source signal acquisition, its parametric representation comprising source position, directional and diffuse signal component estimation, and perceptually sensitive spatial reproduction at the receiver, are the three components of the SSC system. Determining the spatial attributes at the transmitter requires a specialized multi-channel recording facility. A compact microphone array has limited spatial coverage, and it is not effective in estimating the near field source positions. Therefore, we consider a distributed arrangement of compact microphone array nodes for signal acquisition as shown in Fig. \ref{fig:txrxSetup}, which provides for the transmitter to be hands-free, similar to natural acoustic communication. It also improves spatial coverage, and allows for source position estimation. However, the estimation of source spatial attributes using the microphone signals from all the array nodes requires aggregation of node signals at a fusion node, which has additional communication requirements. To avoid this, we consider a distributed computation scheme. The reverberant signal is decomposed into direct and diffuse signal components, and the source direction of arrival (DoA) is computed separately at each node. We use a modified version of the joint spatial filtering and multi-channel linear prediction (MCLP) approach, proposed in \cite{chetupalli2019joint}, for this purpose. The DoA estimates, computed at each of the nodes, are then fused in a geometric formulation to estimate the source position (relative to the microphone arrangement), similar to the approach presented in \cite{Ajdler2004Acoustic}. Finally, the signal components computed at the node nearest to the estimated source position are chosen for transmission to the receiver for spatial reconstruction.
\par The spatial reconstruction at the receiver is with respect to a chosen VL position in the transmitter enclosure; this requires a computation of the source direction relative to the VL position and correction for the amplitude scaling of the direct component since the transmitter signals are estimated at one of the nodes. The diffuse component can be assumed to be similar throughout the enclosure, i.e., a homogenous sound field, with the PSD of the diffuse component being the same at all positions in the enclosure \cite{kuttruff2016room}. 
\par Different spatial audio reproduction schemes are possible at the receiver. For loudspeaker (LS) listening, multiple loudspeakers arranged in a $2$D or $3$D configuration are used to reconstruct spatial speech. In this scheme, the listener position is fixed to be at the center of the LS configuration. In the present work, we consider a simple four loudspeaker setup in a $2$D configuration for reproduction and also assume no additional reverberation at the receiver enclosure. Amplitude panning using vector-base amplitude panning (VBAP) method \cite{pulkki1997virtual} is used to render the direct component, and the diffuse part is rendered equally from all loudspeakers after decorrelation and energy normalization \cite{Kendall1995Decorrelation}. In practice, the reconstructed spatial audio interacts with the receiver location acoustic characteristics, which need to be compensated before LS rendering. Binaural head-phone listening avoids the issue of receiver enclosure reverberation effects. It also has the advantage of minimum setup and the mobility of the listener at the receiver location. However, for effective binaural presentation, the head-related transfer function (HRTF) of the listener is required. Also, the listener should be aware that reproduced spatial scene moves with head rotation. This problem can be solved using a headset with motion tracking, similar to the devices used for virtual reality (VR) applications, and then adapt the rendering of transmitter source location suitably.  
In the present work, we use the method presented in \cite{laitinen2009binaural} for binaural reproduction of spatial speech.
\subsection{Related works}
Spatial speech using the parametric spatial audio methods has been studied in the context of speech teleconferencing systems \cite{pulkki2007spatial, Alexandridis2017Parametric, Parametric2015Kowalczyk}. In directional audio coding (DirAC) \cite{pulkki2007spatial}, signal acquisition using a B-format microphone \cite{gerzon1973periphony} is assumed, and energy analysis of the pressure gradient of recorded signals is used to compute the diffuseness and the source DoA parameters. Single-channel filters are then derived using the estimated parameters and applied to the omnidirectional microphone channel to estimate the direct and diffuse signal components. Although there is higher realism of the recorded sound using a B-format microphone, in particular for audio-visual applications, the fixed location of the 3D microphone becomes a limitation. Hence the parameter estimation accuracy decreases with source distance in reverberant enclosures. DirAC scheme has been extended by using signal acquisition through microphone arrays of omnidirectional elements \cite{keuch2008directional, thiergart2014informed, thiergart2013geometry}. In \cite{keuch2008directional}, microphone signals recorded using a four-channel uniform circular array are used to compute ``pseudo'' B-format microphone signals, followed by DirAC analysis for parameter estimation. In \cite{thiergart2014informed, thiergart2014extracting}, the authors assume plane wave propagation for the directional components and derive multi-channel spatial filters for the estimation of directional and diffuse components. The linearly constrained minimum variance design criterion is used for filter design, and the source DoA is computed using the ESPRIT approach \cite{paulraj1985estimation}. Spatial audio analysis using ad-hoc distributed microphone arrays for signal acquisition has received less attention in the literature. Beamforming based approach of \cite{thiergart2014informed} is considered for signal estimation separately at each node, in \cite{thiergart2013geometry}. The source position (instead of source DoA) is computed by fusion of DoA, estimated separately at each microphone array node using the ESPRIT approach \cite{paulraj1985estimation}. Loudspeaker rendering is considered in most traditional spatial speech/audio approaches; however, the parameters computed in the parametric spatial audio methods can be used for binaural rendering cases using measured HRTFs also. A method for the binaural rendering of signals derived using DirAC spatial analysis is presented in \cite{laitinen2009binaural}.

\section{Transmitter spatial analysis}\label{sec:spatialAnalysis}
	\subsection{Spatial analysis scheme}
	Consider a recording setup with $P$ nodes of compact microphone arrays distributed spatially in the enclosure, and a source is at position $\bld{s}$, and let the chosen virtual listener (VL) position be $\bs{\ell}$. Let $M_p > 1$ be the number of microphones at $p^{th}$ node, and $x_{m,p}[t]$ denote the signal recorded at the $m^{th}$ microphone of the $p^{th}$ node. Therefore
\begin{equation}
x_{m,p}[t] = h_{m,p}[t] \circledast s[t],
\end{equation}
where $h_{m,p}[t]$ is the RIR for the $m^{th}$ microphone position of the $p^{th}$ node. We consider a decomposition of the microphone signal into its directional and diffuse components, $d_{m,p}[t]$ and $r_{m,p}[t]$ respectively,
\begin{equation}
	x_{m,p}[t] = d_{m,p}[t] + r_{m,p}[t],
\end{equation}
where $d_{m,p}[t]$ is due to the direct component and $r_{m,p}[t]$ is due to the reflection components of RIR.
\par Also, let $x_{\bs{\ell}}[t],d_{\bs{\ell}}[t],r_{\bs{\ell}}[t]$ denote the reverberant signal and the constituent components at the VL position $\bs{\ell}$,
\begin{equation}
x_{\bs{\ell}}[t] = d_{\bs{\ell}}[t] + r_{\bs{\ell}}[t].
\end{equation} 
The goal of spatial analysis for SSC is to estimate the signal components $d_{\bs{\ell}}[t], r_{\bs{\ell}}[t]$, and the source position $\bld{s}$. The scheme is to determine the listener position signal decomposition and the source position relative to $\bs{\ell}$ to be transmitted to the receiver. 
\par In a spatially homogeneous sound field, the diffuse component PSD is approximately constant at any position in the enclosure \cite{kuttruff2016room}, i.e.,
\begin{equation}
	\Expectation{\left| \mathcal{F}\inCurly{r_{(.)}[t]}(\omega) \right|^2 } \approx Constant.
\end{equation}
Since $r_{\bs{\ell}}[t]$ is stochastic in nature, the component $r_{m,p}[t]$ at any microphone can be used as an estimate of $r_{\bs{\ell}}[t]$. The direct component at any position in the enclosure is a delayed and attenuated copy of the source signal. Using the position coordinates of the source $\bld{s}$ and the microphone $\bld{m}_{m,p}$, we can approximate:
\begin{eqnarray}
	d_{m,p}[t] \propto \frac{1}{\| \bld{s} - \bld{m}_{m,p} \|_2} s[t-\tau_{m,p}],\\\nonumber ~\mbox{and}~
	d_{\bs{\ell}}[t] \propto \frac{1}{\| \bld{s} - \bs{\ell} \|_2} s[t-\tau_{\bs{\ell}}].
\end{eqnarray}
Thus, the components $d_{\bs{\ell}}[t]$ and $d_{m,p}[t]$ differ mainly in time alignment and amplitude scaling. Properly scaled $d_{m,p}[t]$ can be used as an estimate for $d_{\bs{\ell}}[t]$, accepting small delay differences.
\par
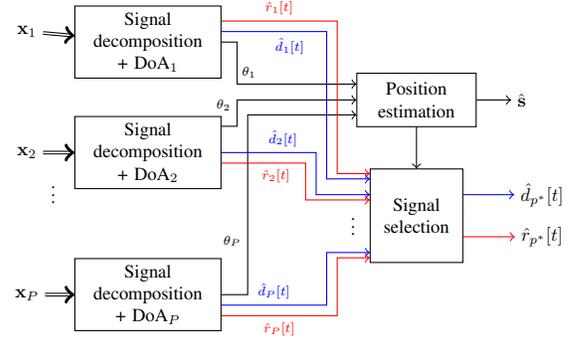
\begin{figure}[t]
	\centering
	\begin{tikzpicture}[scale=0.7,every node/.style={scale=0.7}]
	\node[draw, minimum height=0.4in, text width =1in, align=center] (n1) at (-1.9,0.3) {Signal decomposition + $\mbox{DoA}_1$};
	\node[draw, minimum height=0.4in, text width =1in, align=center] (n2) at (-1.9,-1.8) {Signal decomposition + $\mbox{DoA}_2$};
	\node[] at (-3.7,-2.5) {\vdots};
	\node[draw, minimum height=0.4in, text width =1in, align=center] (nP) at (-1.9,-4.5) {Signal decomposition + $\mbox{DoA}_P$};				
	
	\node[] (in1) at (-4.2,0.5) {$\bld{x}_1$};
	\node[] (in2) at (-4.2,-1.8) {$\bld{x}_2$};
	\node[] (inP) at (-4.2,-4.5) {$\bld{x}_P$};		
	
	\draw[->,double] (in1) -- (n1);\draw[->,double] (in2) -- (n2);\draw[->,double] (inP) -- (nP);
	
	\node[draw, minimum height=0.4in, text width =0.8in, align=center] (pe) at (3.2,-0.8) {Position estimation};
	
	\draw[->] (n1.east) -- ([xshift=3mm]n1.east) |-node[above,pos=0.55]{\scriptsize $\theta_1$} ([yshift=9mm]pe);
	
	\draw[->] ([yshift=5mm]n2.east) -- ([xshift=3mm,yshift=5mm]n2.east) |-node[left,pos=0.4]{\scriptsize $\theta_2$} ([yshift=0mm]pe);		
	
	\draw[->] (nP.east) -- ([xshift=5mm]nP.east) |-node[left,pos=0.15]{\scriptsize $\theta_P$} ([yshift=-8mm]pe);				
	
	\node[draw, minimum height=0.7in, text width =0.6in, align=center] (ss) at (3.2,-3.0) {Signal selection};		
	
	\draw[->] (pe) -- (ss);
	
	\draw[->,blue] ([yshift=2mm] n1.east) --node[below,pos=0.65]{\scriptsize $\hat{d}_{1}[t]$} ([yshift=2mm,xshift=20mm] n1.east) |- ([yshift=7mm]ss.west);
	\draw[->,blue] ([yshift=0mm] n2.east) --node[above,pos=0.65]{\scriptsize $\hat{d}_{2}[t]$} ([yshift=0mm,xshift=18mm] n2.east) |- ([yshift=4mm]ss.west);		
	\draw[->,blue] ([yshift=-2mm] nP.east) --node[above]{\scriptsize $\hat{d}_{P}[t]$} ([yshift=-2mm,xshift=20mm] nP.east) |- ([yshift=-7mm]ss.west);	
	
	\draw[->,red] ([yshift=4mm] n1.east) --node[above]{\scriptsize $\hat{r}_{1}[t]$} ([yshift=4mm,xshift=22mm] n1.east) |- ([yshift=8mm]ss.west);
	\draw[->,red] ([yshift=-2mm] n2.east) --node[below,pos=0.65]{\scriptsize $\hat{r}_{2}[t]$} ([yshift=-2mm,xshift=16mm] n2.east) |- ([yshift=3mm]ss.west);		
	\draw[->,red] ([yshift=-4mm] nP.east) --node[below]{\scriptsize $\hat{r}_{P}[t]$} ([yshift=-4mm,xshift=22mm] nP.east) |- ([yshift=-8mm]ss.west);					
	
	\node[] at (2,-3.1) {\vdots};
	
	\node[] (pos) at (5.2,-0.8) {$\hat{\bld{s}}$};
	\draw[->] (pe) -- (pos);
	
	\node[] (dir) at ([yshift=4mm,xshift=15mm]ss.east) {$\hat{d}_{p^*}[t]$};
	\draw[->,blue] ([yshift=4mm]ss.east) -- (dir.west);
	
	\node[] (de) at ([yshift=-4mm,xshift=15mm]ss.east) {$\hat{r}_{p^*}[t]$};
	\draw[->,red] ([yshift=-4mm]ss.east) -- (de.west);	
	
	\end{tikzpicture}
	\vspace{-5pt}
	\caption{Block diagram description of the proposed spatial analysis scheme.}\label{fig:bd}
\end{figure}

 We first consider the estimation of the signal components $d_p[t],r_p[t]$, and the source DoA $\theta_p$ for each node $p$. We can then select a reference node $p^*$ to select the best signals for transmission to the receiver. The scheme for the estimation of signal components and the source DoA is shown in Fig. \ref{fig:bd}. We modify the joint spatial filtering and MCLP scheme proposed in \cite{chetupalli2019joint} to estimate the direct and diffuse components as well as the source DoA. Thus, we have one set of estimates for each node: $\inCurly{ \hat{d}_p[t], \hat{r}_p[t], \hat{\bs{\theta}}_p, ~1 \leq p \leq P}$, obtained using the above approach. DoA measurements from all the nodes are then used to estimate the source position $\hat{\bld{s}}$; we can then choose node $p^*$ that is nearest to $\hat{\bld{s}}$ and the signal components estimated at the node $p^*$ are used for transmission to the receiver.
	\subsection{Reverberant signal model}
		Let $x_{m}[n,k]$ denote the STFT representation of signal $x_m[t]$  recorded at the $m^{th}$ microphone of $p^{th}$ node (index omitted for brevity), and $n,k$ denote the discrete time and frequency indices respectively, $0\leq k \leq K/2$ where $K$ is the size of discrete Fourier transform (DFT) used in the computation of STFT. We consider delayed multi-channel linear prediction (MCLP) model to estimate the diffuse component of the microphone signal \cite{nakatani2010speech}:
\begin{equation}\label{eqn:mclpRelation}
	x_m[n,k] = \underbrace{\sum\limits_{m'=1}^{M} \sum\limits_{l=0}^{L-1} g_{m,m',k}^*\inSqBrackets{l} x_{m'}[n-D-l,k]}_{r_m[n,k]} + d_m[n,k].
\end{equation}
The delay $D$ is judiciously chosen to avoid over-estimation of the diffuse component, yet preserve natural temporal correlations in speech STFT; the prediction order $L$ depends on the placement of source, microphone, and the length of RIR in the time domain (hence reverberation time of the enclosure). From our previous work \cite{chetupalli2019late}, we choose $D=2$, i,e., prediction using $L$ samples after skipping the immediate past sample, which is found to be optimum. $D=2$ corresponds to $16$~ms delay for STFT computed using $32$~ms frame size with $75\%$ overlap between successive frames; this preserves the direct component and a few early reflections in the residual component $d_m[n,k]$, and the predicted out component $r_m[n,k]$ is the diffuse component.
\par Let $\bld{x}[n,k]= \inSqBrackets{x_1[n,k],\dots,x_M[n,k] }^T$ be the vector of $M$ number of microphone observations, 
\begin{equation}
	\bld{x}[n,k] = \bld{d}[n,k] + \bld{r}[n,k],
\end{equation}
where $\bld{d}[n,k]$ and $\bld{r}[n,k]$ are defined similar to $\bld{x}[n,k]$. The prediction relation in \eqref{eqn:mclpRelation} can be written compactly in vector form as,
\begin{equation}
	\bld{r}[n,k] = \bld{G}_k^H \bs{\phi}_D[n,k],
\end{equation}
where $\bld{G}_k = \left[\bld{g}_{1,k},\dots, \bld{g}_{M,k}  \right]$ is the stack of $M.L \times M$ MCLP coefficients, $\bld{g}_{m,k}$ is the vector of stacked MCLP coefficients 
\begin{multline}
	\bld{g}_{m,k} = \inSqBrackets{ g_{m,1,k}[0],\dots,g_{m,1,k}[L-1],\dots, \right.\\\left. \dots, g_{m,M,k}[0],\dots,g_{m,M,k}[L-1]}^T,
\end{multline}
and $\bs{\phi}_D[n,k]$ is the vector of delayed STFT prediction samples
\begin{multline}
	\bs{\phi}_D[n,k] = \inSqBrackets{x_{1}[n-D,k],\dots,x_{1}[n-D-L+1,k], \right.\\\left. \dots, x_{M}[n-D,k],\dots,x_{M}[n-D-L+1,k]}^T.
\end{multline}
\par We approximate the prediction residual component $\bld{d}[n,k]$ based on the source DoA, using the plane wave signal model,
\begin{gather}
	\bld{d}[n,k] \approx \bld{a}_k d_1[n,k],\nonumber\\
	\bld{a}_k=\inSqBrackets{1,e^{-j2\pi kf_s \tau_{12}/K},\dots,e^{-j2\pi kf_s \tau_{1M}/K}}^T,	
\end{gather}
where $\inCurly{\tau_{12},\dots,\tau_{1M}}$ denote the time-difference-of-arrival (TDOA) of the direct component at each of the microphones with respect to the first (reference) microphone, and $f_s$ is the signal sampling rate. The total signal model, referred to as ``directional MCLP''(dMCLP) from here onwards, is given by
\begin{equation}\label{eqn:sigdecomp}
	\bld{x}[n,k] \approx \bld{a}_k d_1[n,k] + \bld{G}_k^H \bs{\phi}_D[n,k].
\end{equation}
	\subsection{Directional and diffuse component estimation}
		Towards a Bayesian estimation approach, a time-varying complex Gaussian source model is assumed for the signal $d_1[n,k]$,
\begin{equation}
	d_1[n,k] \sim \mathcal{N}_c \inBrackets{0, \gamma_{nk}^{-1}},
\end{equation}
along with distortionless response constraint $\bld{w}_k^H \bld{a}_k = 1$ in the source direction $\bld{a}_k$. $\gamma_{nk}^{-1}$ is the short time power spectral density (PSD) of the signal for the time-frequency bin $(n,k)$.
From \eqref{eqn:sigdecomp}, we have
\begin{equation}
	d_1[n,k] = \bld{w}_k^H \inBrackets{ \bld{x}[n,k] - \bld{G}_k^H \bs{\phi}_D[n,k] }.
\end{equation}
Thus, the model for microphone array observations is,
\begin{multline}\label{eqn:sigmodel11}
	\mathbb{P}\inBrackets{ \bld{x}[n,k] \given \gamma_{nk},\bld{G}_k,\bld{w}_k } =\\ \frac{\gamma_{nk}}{\pi} \exp{\inBrackets{-\gamma_{nk} \left| \bld{w}_k^H \inBrackets{ \bld{x}[n,k] - \bld{G}_k^H \bs{\phi}_D[n,k] } \right|^2 }}.
\end{multline}
Assuming the STFT coefficients $d_1[n,k]$ to be independent across time and frequency (given $\gamma_{nk}$) in the stochastic model of \eqref{eqn:sigmodel11}, we can write
\begin{equation}
	\mathbb{P}\inBrackets{ \bld{X} \given \bs{\Theta}  } = \prod\limits_{n=0}^{N-1} \prod\limits_{k=0}^{K/2}  \mathbb{P}\inBrackets{ \bld{x}[n,k] | \gamma_{nk},\bld{G}_k, \bld{w}_k },
\end{equation}
where  $\bs{\theta}_k=\inCurly{\bs{\gamma}_k=\inCurly{\gamma_{nk},~\forall n},\bld{G}_k,\bld{w}_k}$, and $\bs{\Theta} = \inCurly{\bs{\theta}_k,~\forall k}$ is the set of all parameters. Maximum likelihood (ML) criterion can be used for the filter parameter estimation; i.e.
\begin{equation}
	\mbox{maximize}~ \inCurly{\log \mathbb{P}\inBrackets{ \bld{X} \given \bs{\Theta}  } \triangleq \mathcal{L}\inBrackets{\bs{\Theta}} }.
\end{equation}
Due to the frequency independence assumption, it can be shown that the likelihood function is separable in $k$, and ML estimation requires maximization of,
\begin{multline}\label{eqn:objThetak}
	\mathcal{L} \inBrackets{ \bs{\theta}_k } = \inCurly{- \sum\limits_{n=0}^{N-1} \gamma_{nk} \left| \bld{w}_k^H \inBrackets{ \bld{x}[n,k] - \bld{G}_k^H \bs{\phi}_D[n,k] } \right|^2 \right.\\\left. 
	+ \sum\limits_{n=0}^{N-1} \log \gamma_{nk}},~\mbox{subject to}~ \bld{w}_k^H\bld{a}_k=1,~\forall~k.
\end{multline}
\par We use an iterative approach to maximize the objective function $\mathcal{L} \inBrackets{ \bs{\theta}_k }$. Using the estimates $\hat{\gamma}_{nk}$, and $\hat{\bld{w}}_k$ from the previous iteration, the prediction filter is computed by minimizing the negative likelihood:
\begin{equation}
	-\mathcal{L}\inBrackets{\bld{G}_k} = \sum\limits_{n=0}^{N-1} { \hat{\gamma}_{nk} \left| \hat{\bld{w}}_k^H \inBrackets{ \bld{x}[n,k] - \bld{G}_k^H \bs{\phi}_D[n,k] } \right|^2 }.
\end{equation}
An approximate solution for $\bld{G}_k$ is obtained as in \cite{chetupalli2019joint}:
\begin{gather}
	\label{eqn:gEst}
	\hat{\bld{G}}_k = \bld{R}_{\phi \phi}[k]^{-1} \bld{R}_{\phi x}[k],
\end{gather}
\begin{gather}	
	\bld{R}_{\phi \phi}[k] = \sum\limits_{n=0}^{N-1} \hat{\gamma}_{nk} \bs{\phi}_D[n,k] \bs{\phi}_D[n,k]^H, \\\nonumber
	\mbox{and}~\bld{R}_{\phi x}[k] = \sum\limits_{n=0}^{N-1} \hat{\gamma}_{nk} \bs{\phi}_D[n,k] \bld{x}[n,k]^H.
\end{gather}
The prediction residual is then computed by $\hat{\bld{d}}[n,k] = \bld{x}[n,k] - \hat{\bld{G}}_k^H \bs{\phi}_D[n,k]$. The optimization criterion for $\bld{w}_k$ can be stated using \eqref{eqn:objThetak} as,
\begin{equation}
	\mbox{minimize}~\inCurly{ \bld{w}_k^H  { \bld{R}_{\hat{d}\hat{d}} } \bld{w}_k } ,~\mbox{subject to }\bld{w}_k^H\bld{a}_k=1,
\end{equation}
where $\bld{R}_{\hat{d}\hat{d}} \triangleq \sum\limits_{n=0}^{N-1} \hat{\gamma}_{nk} \hat{\bld{d}}[n,k]\hat{\bld{d}}[n,k]^H$. This optimization problem is similar to the minimum variance distortionless response (MVDR) beamformer design \cite{cox1987robust,habets2010new}. As in MVDR, the objective function is sensitive to errors in the steering vector $\bld{a}_k$, and hence a modified objective in terms of the predicted diffuse component $\hat{\bld{r}}[n,k] = \hat{\bld{G}}_k^H \bs{\phi}_D[n,k]$ is used, whose solution is
\begin{equation}\label{eqn:computewk}
	\hat{\bld{w}}_k = \frac{ \bld{R}_{\hat{r}\hat{r}}^{-1} \bld{a}_k }{ \bld{a}_k^H \bld{r}_{\hat{d}\hat{r}}^{-1} \bld{a}_k},
\end{equation}
where $\bld{R}_{\hat{r}\hat{r}}$ is defined similar to $\bld{R}_{\hat{d}\hat{d}}$. Next, the desired signal variance parameter is computed by maximizing, 
\begin{equation}
	\mathcal{L}\inBrackets{\gamma_{nk}} = \log \gamma_{nk} - {\gamma}_{nk} \left| \hat{\bld{w}}_k^H \hat{\bld{d}}[n,k] \right|^2,
\end{equation}
whose solution is the short-time PSD of the spatial filtered signal $\hat{d}_1[n,k]=\hat{\bld{w}}_k^H \hat{\bld{d}}[n,k]$.
\begin{equation}\label{eqn:gammaEst}
	{\hat{\gamma}}_{nk}^{-1} = f\inBrackets{\left| \hat{d}_1[n,k] \right|^2}.
\end{equation}
Time-domain AR modeling (order $21$) of the signal at time frame $n$ is used to estimate 
the PSD.
\par The direct component is obtained using the estimated MCLP and spatial filters as,
\begin{equation}
	\hat{{d}}_1[n,k] = \hat{\bld{w}}_k^H \inBrackets{ \bld{x}[n,k] - \hat{\bld{G}}_k^H \bs{\phi}_D[n,k] },
\end{equation}
and the diffuse component is obtained as $\hat{{r}}_1[n,k] = x_1[n,k] - \hat{{d}}_1[n,k]$. Since all the estimated quantities are in STFT domain, the overlapped analysis provides for reconstruction of the time domain signals $\hat{{d}}_1[t]$ and $\hat{{r}}_1[t]$. 
\par In \eqref{eqn:computewk}, the computation of spatial filter $\bld{w}_k$ requires the source steering vector $\bld{a}_k$. We consider the SRP-PHAT method \cite{dibiase2000high} to estimate the source direction and the corresponding steering vector. SRP-PHAT is performed on the estimated multichannel prediction residual $\hat{\bld{d}}[n,k]$, which comprises of stronger direct component compared to the microphone signals $\bld{x}[n,k]$. Let $\bld{a}_{k,\theta_l}$ denote the steering vector for a far-field source at an angle $\theta_l$ at frequency bin $k$, 
\begin{equation}
	\bld{a}_{k,\theta_l} = \inSqBrackets{1~e^{\inBrackets{\frac{-j2\pi kf_s \tau_{21}(\theta_l)}{K}}} \dots e^{\inBrackets{\frac{-j2\pi kf_s \tau_{M1}(\theta_l)}{K}}}}^T,
\end{equation}
where $\tau_{m1}(\theta_l)$ is the TDOA between the $m^{th}$ microphone and the first microphone, for a source in the direction of $\theta_l$. In SRP-PHAT \cite{dibiase2000high}, the direction of the source is estimated by solving,
\begin{equation}
	\hat{\theta}[n] = \underset{\theta_l}{\arg\max}~\sum\limits_{k=0}^{K/2} \left| \bld{a}_{k,\theta_l}^H \hat{\bld{d}}_f[n,k] \right|^2,
\end{equation}
where $\hat{\bld{d}}_f[n,k] = \hat{\bld{d}}[n,k]/\left| \hat{\bld{d}}[n,k] \right|$ is the PHAT filtered (phase only) signal \cite{knapp1976generalized}. We compute the DoA estimate for several consecutive frames and then choose the mode of frame-wise DoAs as the estimated DoA $\hat{\theta}$ for the stationary source:
\begin{equation}\label{eqn:aEst}
	\hat{\theta} = \mbox{mode}\inBrackets{ \inCurly{\hat{\theta}[n],~\forall~n} },~\mbox{and}~\hat{\bld{a}}_k = \bld{a}_{k,\hat{\theta}}.
\end{equation}
\par Alternatively, GCC-PHAT \cite{knapp1976generalized,carter1973smoothed} based method can be used to compute the TDOA between microphone pairs, followed by source DoA estimation. For the case of two microphones in the array, SRP-PHAT and GCC-PHAT provide similar results. But the correlation computed on the integer delay grid in GCC-PHAT limits the DoA accuracy, and requires inter sample interpolation to improve the accuracy, hence SRP-PHAT is preferred.
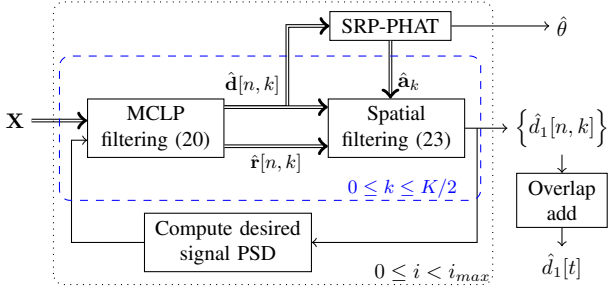
\begin{figure}[t]
	\centering
	\begin{tikzpicture}[scale=0.8, every node/.style={scale=0.8}]
	\node[] (inNode) at (-1.0in,0.05in) {${\bf X}$}; 
	\node[draw,text width=0.8in,align=center] (filter) at (-0.2,0) {MCLP filtering (20)};		
	\node[draw,text width=0.7in,align=center] (filtest) at (3.7,1.7) {SRP-PHAT};		
	\draw[double,->] ([yshift=3.5mm] filter.east) -| node[left,pos=0.65]{\small $\hat{\bld{d}}[n,k]$} ([xshift=-7mm] filtest.west) --  (filtest.west);		
	\node[draw,text width=0.8in,align=center] (residual) at (3.8,0) {Spatial filtering (23)};		
	\draw[double,->] (filtest) --node[above,right,pos=0.7]{$\hat{\bld{a}}_k$} (filtest |- residual.north);
	\node[draw,text width=1.0in,align=center] (psd) at (1,-1.9) {Compute desired signal PSD};
	
	\draw[double,->] ([yshift=3.5mm]filter.east) --  ([yshift=3.5mm]residual.west);		
	\draw[double,->] ([yshift=-3mm]filter.east) --node[below]{\small $\hat{\bld{r}}[n,k]$}  ([yshift=-3mm]residual.west);
	
	\draw[->] ([xshift=2mm]residual.east) |- (psd.east);
	\draw[->] (psd.west) -- (-1.6,-1.9) |- ([yshift=-2mm]filter.west);
	\draw[double,->] (inNode) -- (inNode -| filter.west);
	\node[] (outNode) at (6.55,0) {$\inCurly{\hat{d}_1[n,k]}$};

	\node[text width=0.5in,align=center,draw] (ola) at (6.55,-1.2) {Overlap add};
	\node[] (outNode2) at (6.55,-2.3) {$\hat{d}_1[t]$};
	\draw[->] (outNode) -- (ola);\draw[->] (ola) -- (outNode2);
	
	\draw[dashed, rounded corners,blue] (-1.8,-1.2) rectangle (5.2,1.2);
	\draw[dotted, rounded corners] (-1.9,-2.6) rectangle (5.4,2.1);
	\node[] at (4.4,-2.4) {$0\leq i < i_{max}$};
	\node[blue] at (3.9,-1.0) {\small $0 \leq k \leq K/2$};
	\draw[->] (residual.east) -- (outNode);		
	
	\node[] (outAngle) at (6.55,1.7) {$\hat{\theta}$};
	\draw[->] (filtest) -- (outAngle);
	\end{tikzpicture}
	\vspace{-5pt}
	\caption{Iterative directional MCLP estimation scheme.}\label{fig:dirMCLPBd}
\end{figure}
\par A block diagram of the dMCLP scheme for the joint reverberant signal decomposition and DoA estimation is shown in Fig. \ref{fig:dirMCLPBd}. The iterative scheme consists of a cascade of MCLP and spatial filtering stages. At iteration-$i$, the microphone signals and the estimate of direct component signal PSD from the iteration $i-1$ are used to compute the MCLP coefficients for each frequency bin $k$, using \eqref{eqn:gEst}. The prediction residual is then used to estimate the source DoA $\hat{\theta}$ using SRP-PHAT method. The steering vector $\hat{\bld{a}}_k$ is computed for each $k$, as in \eqref{eqn:aEst}, and used to design the spatial filter of \eqref{eqn:computewk}. Direct component signal is estimated as the output of spatial filter, which is then used to compute the PSD estimate for next iteration. The estimate of $\hat{\gamma}_{nk} = \|\bld{x}[n,k]\|^2/M$ is used as the initialization for the first iteration, and the algorithm is executed for a pre-fixed number of iterations $i_{max}$.
	\subsection{Source position}
		Source DoAs, estimated at each node, $\inCurly{\hat{\theta}_p,~1 \leq p \leq P}$, are computed using SRP-PHAT through the iterations of dMCLP discussed above. We then compute the source position by the intersection of lines representing source direction from each of the nodes. For the purpose of discussion below, we assume the source and microphones to be in the same azimuth plane and the position estimation is in a two dimensional space. Extension to three dimensional source position is straight forward, using non-coplanar microphone array nodes which can estimate the source elevation angle also.
%
%
\par Let the line $(\bld{n}_p + r_p \bld{i}_p)$, be the contour of possible source positions, where $r_p$ is the unknown distance of the source from $p^{th}$ node, $\bld{n}_p$ is the known position of $p^{th}$ node and $\bld{i}_p$ is the unit vector pointing in the estimated direction $\hat{\theta}_p$, illustrated in Fig. \ref{fig:srcPosEstIllu}. The source position $\bld{s}$ is estimated by solving for the intersection of lines,
\begin{equation}
	\hat{\bld{s}} = \mbox{intersect}\inCurly{\inBrackets{\bld{n}_1 + r_1 \bld{i}_1},\dots,\inBrackets{\bld{n}_P + r_P \bld{i}_P}}.
\end{equation}
A common intersection point may not exist in practice, because of estimation errors in DoA. We choose mean of the intersection points of pairs of DoA lines, computed after removing outliers, as the estimated source position.
	\subsection{Signal selection}
		We consider the node nearest to the estimated source position $\hat{\bld{s}}$ as the best node $p^*$. The directional component $\hat{d}_{1,p^*}[t]$ and the diffuse component $\hat{r}_{1,p^*}[t]$ estimated at the reference microphone $r=1$ at node $p^{*}$ are chosen for transmission to the receiver. Thus, $\hat{d}[t] = \hat{d}_{1,p^*}[t]$ and $\hat{r}[t] = \hat{r}_{1,p^*}[t]$, along with the source position estimate $\hat{\bld{s}}$, are the parameters to be quantized and sent over the channel.
\section{Receiver spatial reconstruction}
	The directional and the diffuse component signals estimated at the transmitter best node are used to compute the LS signals for spatial reconstruction at the receiver. The directional component is assumed to be coming from a point source and hence can be rendered using VBAP \cite{pulkki1997virtual}. However, the diffuse component is decorrelated and hence rendered from all loudspeakers to create the ambiance of the transmitter location.
\begin{figure}[t]
	\centering
	\begin{minipage}{0.48\linewidth}
		\centering
		\begin{tikzpicture}[scale=0.45,every node/.style={scale=0.45}]
		\draw[] (0,0) rectangle (8,8);				
		\speakerP{(3.25,2.5)}{0.4}{red};
		\node[] at (2.45,2.7) {\large $\bld{s}$};
		\mic{(2,0)}{90}{0.4};
		\mic{(2.4,0)}{90}{0.4};
		\draw[rounded corners=0.5mm] (1.8,0.3) rectangle (2.6,0.65);
		\mic{(6,8)}{-90}{0.4};
		\mic{(6.4,8)}{-90}{0.4};
		\draw[rounded corners=0.5mm] (5.8,7.35) rectangle (6.6,7.7);			
		\mic{(0,5.0)}{0}{0.4};
		\mic{(0,5.4)}{0}{0.4};
		\draw[rounded corners=0.5mm] (0.3,4.8) rectangle (0.65,5.6);			
		\mic{(8,2)}{180}{0.4};
		\mic{(8,2.4)}{180}{0.4};
		\draw[rounded corners=0.5mm] (7.35,1.8) rectangle (7.7,2.6);			
		\speakerM{(5.7,4.1)}{0.8}{90};
		\node[] at (6,3.5) {\large $\bs{\ell}$};
		\draw[->,dotted] (5.7,4.1) --node[above,pos=0.8,red]{\large {$\|\bs{\ell}-{\bld{s}}\|$}} (3.3,2.9);			
		\draw[->,dotted] (2.15,0.45) --node[right,pos=0.5]{\large {$\|\bld{n}_{1}-{\bld{s}}\|$}} (2.9,2.5);
		\node[] at (1.5,0.2){\large $\bld{n}_1$};
		\node[rotate=90] at (7.7,3){\large $\bld{n}_2$};
		\node[rotate=180] at (5.4,7.7){\large $\bld{n}_3$};
		\node[rotate=-90] at (0.2,4.4){\large $\bld{n}_4$};
		
		\draw[->,dotted] (5.7,4.1) -- (5.7,6.0);
		\draw[dashed,red] ([shift=(90:1cm)]5.7,4.1) arc (90:205:1cm);
		\node[red] at (4.5,4.5) { $\theta$};
		
		\node[] at (3,0.3){\large $\hat{d}[t]$};
		\node[rotate=90] at (6.1,5){\large ${d}^*[t]$};
		\end{tikzpicture}
		\centerline{(a) Transmitter (Tx)}
	\end{minipage}
	\begin{minipage}{0.48\linewidth}
		\centering
		\begin{tikzpicture}[scale=0.45,every node/.style={scale=0.45}]
		\draw[] (0,0) rectangle (8,8);
		\speakerP{(3.2,6.0)}{0.4}{red,very thin,rotate=0};
		\speakerM{(4,4)}{0.8}{0};
		\speakerF{(0.15,0.1)}{1}{0};
		\speakerF{(7.1,0.1)}{1}{0};
		\speakerF{(7.1,6.9)}{1}{0};
		\speakerF{(0.15,6.9)}{1}{0};
		
		\draw[->,dotted] (4,4) --node[right,pos=0.6]{\large$\bs{\ell}_1$} (4+1.414,4+1.414);
		\draw[->,dotted] (4,4) --node[left,pos=0.6]{\large $\bs{\ell}_2$} (4-1.414,4+1.414);
		\draw[->,dotted] (4,4) --node[left,pos=0.6]{\large $\bs{\ell}_3$} (4-1.414,4-1.414);
		\draw[->,dotted] (4,4) --node[right,pos=0.6]{\large $\bs{\ell}_4$} (4+1.414,4-1.414);			
		
		\draw[->,dotted] (4,4) -- (5.5,4);
		\draw[->,dotted] (4,4) -- (3.0,6);
		\draw[dashed,red] ([shift=(0:1cm)]4,4) arc (0:115:1cm);
		\node[red] at (4.1,5.2) {\large $\theta$};
		\end{tikzpicture}
		\centerline{(b) Receiver (Rx)}
	\end{minipage}
	\caption{Transmitter and receiver setup for spatial speech communication.}\label{fig:txrxSetup2}	
\end{figure}
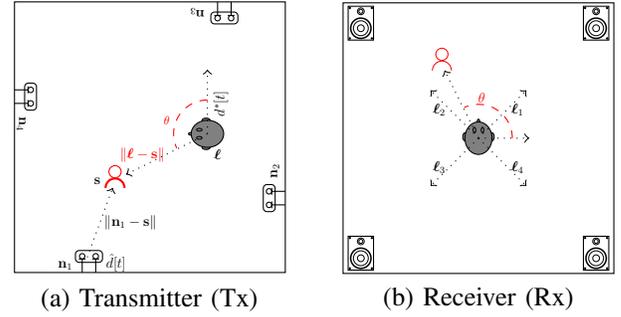
\par Let $\bs{\ell}$ denote the virtual listener (VL) position in the transmitter space, which could be through a selection of coordinates by the receiver, who could move closer to the source or another location for better listening, which can be achieved through a GUI of the transmitter space made available at the receiver. Given the estimated source position $\hat{\bld{s}}$, the angular position of the source with respect to the VL position $\bs{\ell}$ is computed as $\angle{\inBrackets{{\bs{\ell}}-\hat{\bld{s}}}}$. Note that the directional component of the source signal is determined at the node nearest to the estimated source position and may have a different amplitude level compared to the required directional component at the VL position. Hence, to compensate for this, the received directional component signal is scaled by the difference in the radial distances between the node $p^*$ and the VL position $\bs{\ell}$ for the source at $\bld{s}$.
\begin{equation}\label{eqn:signaltransformation}
	{d}^*[t]=\frac{\|\hat{\bld{s}} - {\bld{n}_{p^*}}\|}{\|\hat{\bld{s}} - \bs{\ell}\|} \hat{d}[t],
\end{equation}
where ${\bld{n}_{p^*}}$ is the best node position. We assume the diffuse component to be spatially homogeneous, and hence it is the same at the VL position as well as the reference node, $r^*[t]=\hat{r}[t]$. 
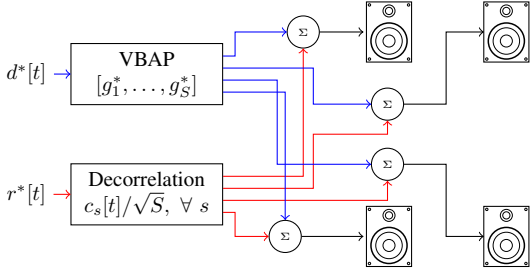
\begin{figure}[t]
	\centering
	\begin{tikzpicture}[scale=0.8,every node/.style={scale=0.8}]
		\centering
		\node[] (d) at (-4.5,2.5) {$d^*[t]$};
		\node[] (r) at (-4.5,0.5) {$r^*[t]$};		
		
		\speakerF{(1.15,-1+0.3)}{1}{0};
		\speakerF{(3.1,-1+0.3)}{1}{0};
		\speakerF{(3.1,3-0.3)}{1}{0};
		\speakerF{(1.15,3-0.3)}{1}{0};		
		
		\node[draw,minimum height=0.4in,text width=0.9in,align=center] (v) at (-2.5,2.5) {VBAP $\inSqBrackets{g_1^*,\dots,g_S^*}$};
		\node[draw,minimum height=0.4in,text width=0.9in,align=center] (D) at (-2.5,0.5) {Decorrelation $c_s[t]/\sqrt{S},~\forall~s$};

		\node[draw,circle] (s1) at (0.1,3.5-0.3) {\tiny $\Sigma$};
		\node[draw,circle] (s2) at (-0.2,-0.5+0.3) {\tiny $\Sigma$};
		\node[draw,circle] (s3) at (1.5,1) {\tiny $\Sigma$};
		\node[draw,circle] (s4) at (1.5,2) {\tiny $\Sigma$};						

		\draw[->,red] ([yshift=-3mm]D.east) -- ([yshift=-3mm,xshift=2mm]D.east) |- (s2);
		\draw[->,red] ([yshift=-1mm]D.east)  -| (s3.south);
		\draw[->,red] ([yshift=1mm]D.east) -- ([yshift=1mm,xshift=15mm]D.east) -- ([yshift=10mm,xshift=15mm]D.east) -| (s4.south);
		\draw[->,red] ([yshift=3mm]D.east) -| (s1.south);
		
		\draw[->,blue] ([yshift=-3mm]v.east) -| (s2.north);
		\draw[->,blue] ([yshift=-1mm]v.east) --([yshift=-1mm,xshift=9mm]v.east) |- (s3.west);
		\draw[->,blue] ([yshift=1mm]v.east) --([yshift=1mm,xshift=15mm]v.east) |- (s4.west);
		\draw[->,blue] ([yshift=3mm]v.east) --([yshift=3mm,xshift=2mm]v.east) |- (s1.west);
		
		\draw[->] (s1) -- (1.1,3.2);
		\draw[->] (s2) -- (1.1,-0.2);
		\draw[->] (s3.east) -- ([xshift=7mm]s3.east) |- (3.1,-0.2);
		\draw[->] (s4.east) -- ([xshift=7mm]s4.east) |- (3.1,3.2);
		
		\draw[->,blue] (d) -- (v);
		\draw[->,red] (r) -- (D);
\end{tikzpicture}
\vspace{-5pt}
\caption{Rx spatial reconstruction scheme for a $S=4$ channel LS rendering.}\label{fig:repscheme}
\vspace{-5pt}
\end{figure}
\par At the receiver, the mono signals $d^*[t]$ and $r^*[t]$ are separately mapped to the LS signals $\inCurly{d_s[t], r_s[t],1\leq s \leq S}$ using the scheme discussed below, and the signals $\inCurly{d_s[t]+r_s[t], ~\forall~s}$ are finally played through the loudspeakers. A block diagram of the receiver reconstruction of spatial source signal is shown in Fig. \ref{fig:repscheme}. 
\subsection{Direction component}
The directional component is reconstructed using the VBAP method \cite{pulkki1997virtual}. In VBAP, each LS is represented by a unit-length direction vector $\{\bld{i}_{s},~\forall s=1,\dots,S\}$ with reference to a fixed listener position, and the direction of source $\bld{s}$ is expressed as,
\begin{equation}
	\bld{s}=\sum\limits_{s=1}^{S} g_s \bld{i}_{s},~g_s\geq 0, \forall~s=1,\dots,S.
\end{equation}
We assume that the loudspeakers and the associated power amplifiers are suitably calibrated and compensated to obtain uniform, comfortable loudness at the listener position. The weights $\{g_s\}$ are normalized to a unit energy,
\begin{equation}
	g_s^*= g_s/E_g,~ E_g \triangleq {\sqrt{\sum\limits_{s=1}^{S} g_s^2}},~\forall~s=1,\dots,S,
\end{equation}
and applied to the directional component to generate the LS signals, ${d}_s[t]=g_s^* {d}^*[t],~\forall~s$. The loudspeakers at the receiver position may not be exactly in a square formation; this can be compensated using the direction vectors $\bld{i}_s$ and the corresponding weights ${g}_s$ for a suitable listener position and the LS positions at the receiver. 
\subsection{Diffuse component}\label{sec:diffuseRender}
The diffuse component is rendered such that the sound is perceived with natural reverberation of the transmit space without the dominant direction of the source but rather enveloping the listener (it is assumed that the receiver listener enclosure is reverb-free). The LS signals are derived from the single-channel signal $r^*[t]$, such that they have the same spectral properties (PSD), but they are not correlated. Deriving the decorrelated signals is important \cite{Kendall1995Decorrelation} to (i) reduce the timbral artifacts such as comb filter effect, (ii) avoid precedence effect \cite{Hans1949Precedence}, and (iii) to create a real diffuse sound. Several approaches exist, such as phase randomization, or passing through suitably designed filters. In the latter approach, a filter is designed such that the magnitude spectrum is a unit constant, but the phase is changed to achieve inter LS decorrelation. The LS signals are computed by first passing $r^*[t]$ through the decorrelation filters $g_s[t]$, one for each LS, followed by energy normalization to maintain the relative levels of direct and diffuse components at the listener position; i.e., $r_s[t] = \inBrackets{r^*[t] \circledast g_s[t]}/\sqrt{S},~S=4$, as at the transmitter. We consider the method given in \cite{pulkki2018firstorder} for the decorrelation filters, which uses sub-band impulses at random time instants in their construction.
\section{Experimental evaluation}
	\subsection{Simulated enclosure Tx space}
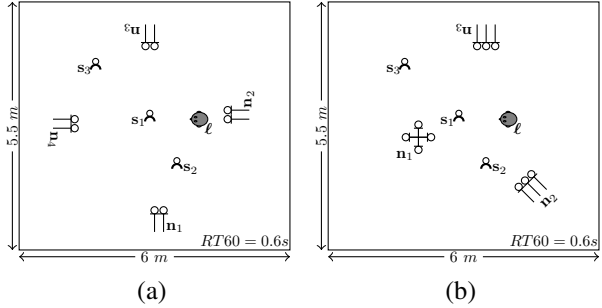
\begin{figure}[t]
	\centering
	\begin{minipage}[b]{0.45\linewidth}
		\begin{tikzpicture}[scale=0.6,every node/.style={scale=0.6}]
		\draw[] (0,0) rectangle (6,5.5);
		
		\mic{(3.2,0.4)}{90}{0.4};
		\mic{(3.0,0.4)}{90}{0.4};\node[] at (3.45,0.5) {$\bld{n}_1$}; 
		
		\mic{(5.1,3.1)}{180}{0.4};
		\mic{(5.1,2.9)}{180}{0.4};\node[rotate=90] at (5.1,3.35) {$\bld{n}_2$}; 		
		
		\mic{(2.8,5)}{270}{0.4};
		\mic{(3.0,5)}{270}{0.4};\node[rotate=180] at (2.5,4.9) {$\bld{n}_3$};		
		
		\mic{(0.75,2.7)}{0}{0.4};
		\mic{(0.75,2.9)}{0}{0.4};\node[rotate=270] at (0.75,2.45) {$\bld{n}_4$};													
		
		\node[] (x1) at (3,-0.15) {$6~m$};		
		\draw[<-] (0,-0.15) -- (x1);\draw[->] (x1) -- (6,-0.15);
		\node[rotate=90] (y1) at (-0.15,2.75) {$5.5~m$};				
		\draw[<-] (-0.15,0) -- (y1);\draw[->] (y1) -- (-0.15,5.5);		
		\node[] at (5.0,0.2) {$RT60=0.6s$};
		
		\speakerP{(3.0,2.85)}{0.15}{};\node[] at (2.65,2.85) {$\bld{s}_1$}; 
		\speakerP{(3.6,1.8)}{0.15}{};\node[] at (3.8,1.8) {$\bld{s}_2$}; 
		\speakerP{(1.8,4.0)}{0.15}{};\node[] at (1.45,4.0) {$\bld{s}_3$}; 				
		\speakerM{(4,2.9)}{0.3}{90};\node[rotate=0] at (4.2,2.7) {$\bs{\ell}$};
		\end{tikzpicture}
		\centerline{(a)}		
	\end{minipage}
	\begin{minipage}[b]{0.45\linewidth}
		\begin{tikzpicture}[scale=0.6,every node/.style={scale=0.6}]
		\draw[] (0,0) rectangle (6,5.5);
		
		\mic{(2,2.3)}{90}{0.4};
		\mic{(2.0,2.7)}{270}{0.4};
		\mic{(1.8,2.5)}{0}{0.4};
		\mic{(2.2,2.5)}{180}{0.4};				
		\node[] at (1.7,2.1) {$\bld{n}_1$};
		
		\mic{(4.7,1.2)}{135}{0.4};
		\mic{(4.841,1.34)}{135}{0.4};
		\mic{(4.56,1.06)}{135}{0.4};
		\node[rotate=45] at (4.9,1.1) {$\bld{n}_2$};
		
		\mic{(3.3,5)}{270}{0.4};
		\mic{(3.7,5)}{270}{0.4};
		\mic{(3.5,5)}{270}{0.4};\node[rotate=180] at (3.0,4.9) {$\bld{n}_3$};
		
		\node[] (x1) at (3,-0.15) {$6~m$};		
		\draw[<-] (0,-0.15) -- (x1);\draw[->] (x1) -- (6,-0.15);
		\node[rotate=90] (y1) at (-0.15,2.75) {$5.5~m$};				
		\draw[<-] (-0.15,0) -- (y1);\draw[->] (y1) -- (-0.15,5.5);		
		\node[] at (5.0,0.2) {$RT60=0.6s$};
		
		\speakerP{(3.0,2.85)}{0.15}{};\node[] at (2.65,2.85) {$\bld{s}_1$}; 
		\speakerP{(3.6,1.8)}{0.15}{};\node[] at (3.8,1.8) {$\bld{s}_2$}; 
		\speakerP{(1.8,4.0)}{0.15}{};\node[] at (1.45,4.0) {$\bld{s}_3$}; 				
		\speakerM{(4,2.9)}{0.3}{90};\node[rotate=0] at (4.2,2.7) {$\bs{\ell}$};
		\end{tikzpicture}
		\centerline{(b)}		
	\end{minipage}
	\vspace{-5pt}
	\caption{Simulated Tx enclosure with reverberation and two configurations of microphone arrays for spatial speech communication, (a) $\bld{C}_1$ with $2$ mics each at $P=4$ nodes, (b) $\bld{C}_2$ with variable number of mics (3,3,4) at $P=3$ nodes. Three unknown source positions $\bld{s}_1,\bld{s}_2,~\mbox{and}~\bld{s}_3$ and the chosen virtual listener position $\bs{\ell}$ are as shown.}\label{fig:recordingSetup}
	\vspace{-5pt}
\end{figure}

We consider two configurations of nodes for the distributed microphone arrays as shown in Fig. \ref{fig:recordingSetup}. The configuration-$\bld{C}_1$, consists of $4$ nodes with $2$ microphones at each node, and an inter microphone spacing of $20$~cm. The transmitter enclosure size is $6 \times 5.5 \times 4.5~\mbox{m}^3$, indicated in Fig. \ref{fig:recordingSetup}(a). The configuration-$\bld{C}_2$, shown in Fig. \ref{fig:recordingSetup}(b), comprises of $3$ nodes, two nodes with $3$~microphones each along a straight line with an inter-microphone spacing of $10$~cm; the third node is a $4$ channel UCA of radius $10$~cm. Reverberation time ($RT60$) of the enclosure is chosen as $0.6$~s to suit the typical acoustically untreated living/office rooms. We consider, three random source positions having azimuth coordinates of $(3,2.85)$~m, $(3.6,1.8)$~m, and $(1.8,4)$~m respectively at an elevation of $1.1$~m. The sources and the microphones are assumed to be in the same azimuth plane at $1.1$~m. The image method \cite{allen1979image} is used to simulate the RIRs between the source and each of the $8~\mbox{or}~10$ microphone positions \cite{rirgenerator}.
\subsection{Tx signal decomposition and position estimation}
First, we study the performance of direct component signal estimation and the source position estimation. Ten clean signals ($5$ each from male and female speakers) of the TIMIT database \cite{timit} are convolved with the simulated RIRs corresponding to each source position and mic position to obtain the microphone signals. STFT of the microphone signals is computed using a Hann window of duration $32$~ms, and $75\%$ overlap between successive analysis windows for the dMCLP analysis. The size of the DFT in STFT is chosen the same as the window size ($512$ for the signal sampling rate of $16~\mbox{KHz}$). dMCLP order $L$ is chosen as $32$, and the delay parameter $D$ is chosen to be $2$; i.e., dMCLP utilizes $L=32$ STFT samples after the immediate past sample for the linear prediction. The dMCLP parameters are kept the same for all the microphones. In SRP-PHAT computation for DoA, we consider a fine search grid of $1^o$ spacing, which corresponds to $<7$~cm resolution in spatial location, for the chosen enclosure size.
\begin{figure}[t]
	\begin{minipage}[b]{0.49\linewidth}
		\centerline{\includegraphics[width=1.6in,height=1.4in]{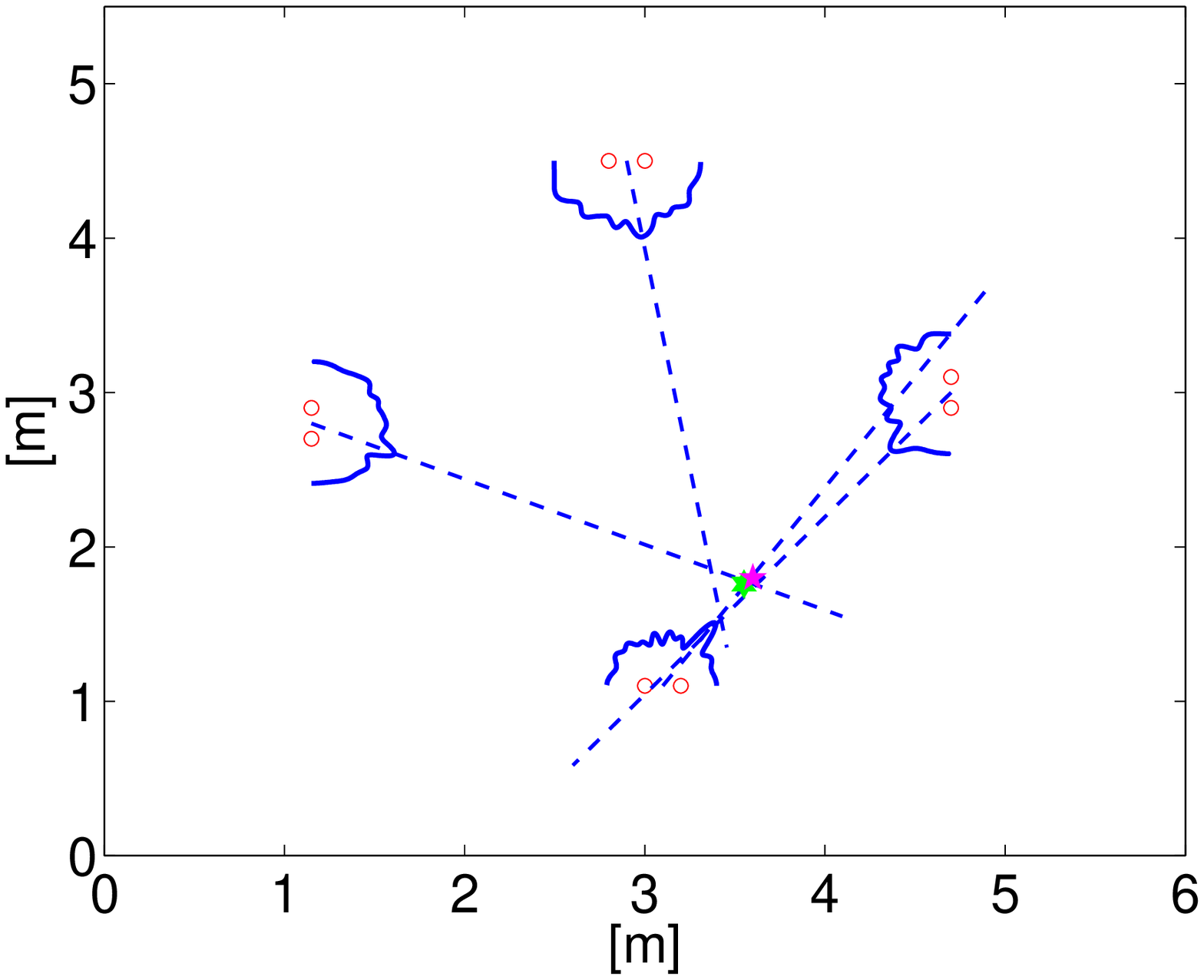}}
		\centerline{(a) $C_1$}
	\end{minipage}
	\begin{minipage}[b]{0.49\linewidth}
		\centerline{\includegraphics[width=1.6in,height=1.4in]{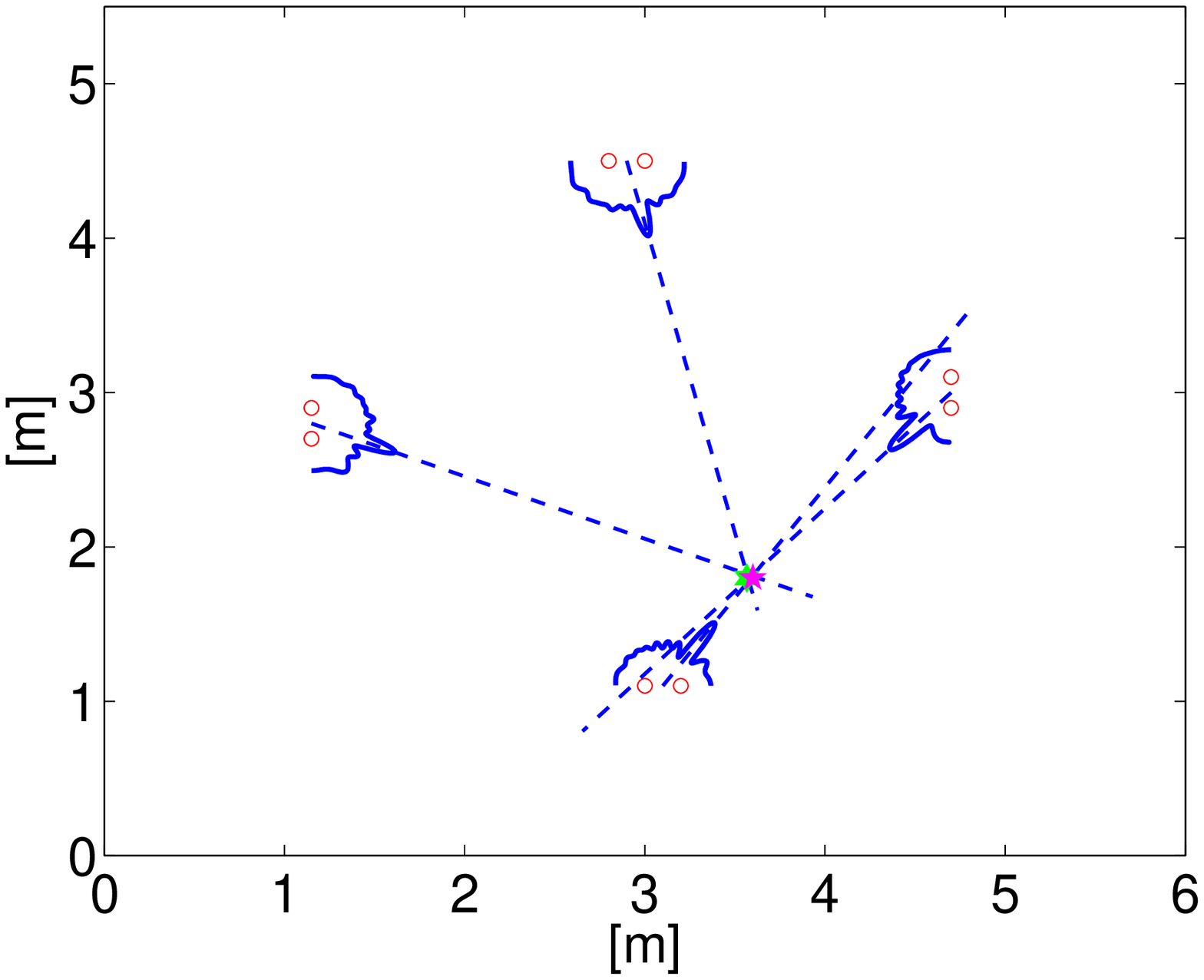}}
		\centerline{(b) $C_1$}
	\end{minipage}
	\begin{minipage}[b]{0.49\linewidth}
		\centerline{\includegraphics[width=1.6in,height=1.4in]{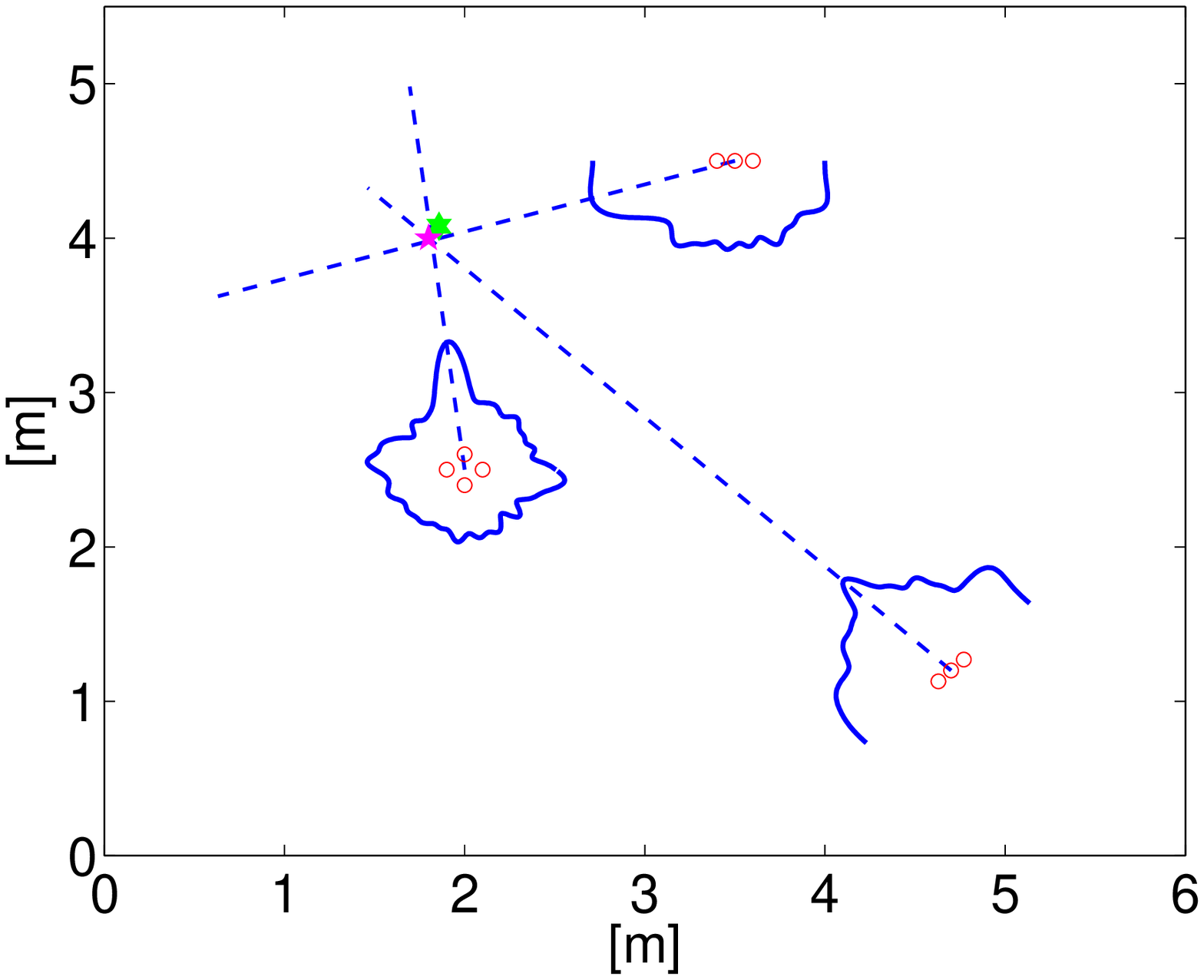}}
		\centerline{(c) $C_2$}
	\end{minipage}
	\begin{minipage}[b]{0.49\linewidth}
		\centerline{\includegraphics[width=1.6in,height=1.4in]{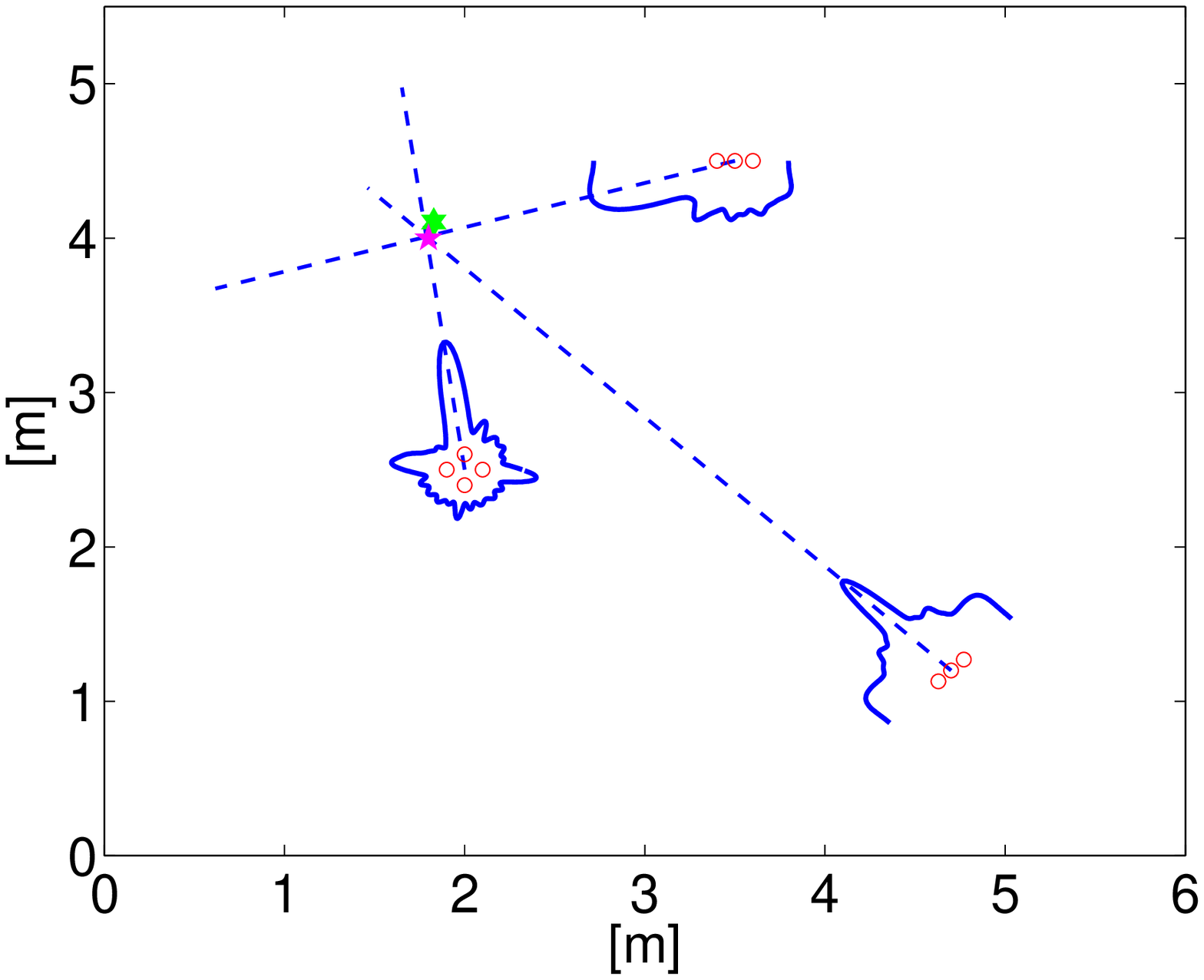}}
		\centerline{(d) $C_2$}
	\end{minipage}	
	\vspace{-5pt}
	\caption{Source position estimation illustration using two different microphone array node configurations. SRP-PHAT directional response contour at each node is shown (a,c) computed using the reverberant microphone signals $\bld{x}[t]$, and (b,d) computed using dereverb signals $\bld{d}[t]$ of MCLP. The lines corresponding to estimated node DoAs are shown along with the estimated source position (green) and ground truth source position (magenta). The SRP-PHAT estimation in (b,d) is significantly better than that of (a,c).}\label{fig:srcPosEstIllu}
	\vspace{-5pt}
\end{figure}
\par An illustration of the position estimation error using the microphone array configuration $\bld{C}_1$ for the speech source at position $\bld{s}_2$ is shown in Fig. \ref{fig:srcPosEstIllu}(b) along with the SRP-PHAT directional response contours for each node. The performance can be compared to the same position estimation using the reverberant microphone signals without the MCLP shown in Fig. \ref{fig:srcPosEstIllu}(a). Clearly, the reverberant signals show more erroneous direction estimates compared to that of dMCLP shown in Fig. \ref{fig:srcPosEstIllu}(b). Similarly, Figs. \ref{fig:srcPosEstIllu}(c,d) illustrate position estimation for the configuration $\bld{C}_2$ and the source position $\bld{s}_3$. Here also we can see narrower beamwidths and better source position estimate in \ref{fig:srcPosEstIllu}(d) compared to \ref{fig:srcPosEstIllu}(c). The better position estimation can be attributed to the suppression of the diffuse component through the dMCLP scheme. (This robustness to directionality through dMCLP becomes more important perceptually in the presence of multiple sources or higher Tx reverberation). Also, the $4$ microphone UCA gives the best direction estimation compared to the other nodes of $2$ or $3$ mics.
\begin{table}[t]
	\centering
	\caption{RMS position error ({RMSPE}) for two configurations $\bld{C}_1,\bld{C}_2$ of microphone array nodes and three source positions in an enclosure of $RT60=0.6~sec$. Absolute DoA error at each node is also shown along with the source position error.}\label{tab:posEstPerfs}	
	\begin{tabular}{|c|c|c|c|c|c|}
		\hline
		\multirow{2}{*}{\shortstack[c]{Src. \\ Pos.}} &  & \multicolumn{2}{c|}{Node Conf. $\bld{C}_1$} & \multicolumn{2}{c|}{Node Conf. $\bld{C}_2$}\\\cline{3-6}
		&  & {Unproc.} & {dMCLP} &  {Unproc.} & {dMCLP} \\ \hline
		\multirow{5}{*}{\shortstack[c]{$\bld{s}_1$}} 
		& \multirow{4}{*}{\shortstack[c]{DoA \\ Error \\($^o$)}} & 1.27 & 0.27 & 0.29 & 0.29 \\
		&  & 1.04 & 0.04 & 0.14 & 0.14\\
		&  & 1.47 & 0.47 & 0.79 & 0.29\\
		&  & 0.55 & 0.55 & - & -\\\cline{2-6}
		& RMSPE (cm)& 0.8  & 1.66 & 5.93 & 5.02\\		
		\hline
		\multirow{5}{*}{\shortstack[c]{$\bld{s}_2$}} 
		& \multirow{4}{*}{\shortstack[c]{DoA \\ Error \\($^o$)}} & 0.54 & 0.54 & 0.37 & 0.37 \\
		&  & 0.51 & 0.31 & 0.39 & 0.39 \\
		&  & 1.33 & 1.53 & 1.12 & 0.12\\
		&  & 0.20 & 0.20 & - & -\\\cline{2-6}
		& RMSPE (cm) & 5.38 & 5.83 & 5.42 & 1.13 \\			
		\hline		
		\multirow{5}{*}{\shortstack[c]{$\bld{s}_3$}} 
		& \multirow{4}{*}{\shortstack[c]{DoA \\ Error \\($^o$)}} & 0.35 & 1.04 & 0.59 & 0.54 \\
		& & 9.13 & 0.13 & 0.01 & 0.01 \\
		& & 0.04 & 0.56 & 0.5 & 0.39 \\
		& & 0.46 & 0.56 & - & - \\\cline{2-6}
		& RMSPE (cm) & 13.9 & 3.50 & 10.14  & 10.72\\
		\hline
	\end{tabular}
	\vspace{-5pt}
\end{table}
\par Table \ref{tab:posEstPerfs} shows the average absolute error in the estimated {DoA} at each node and the RMS error of the source position estimation for the two node configurations $\bld{C}_1,\bld{C}_2$ of microphone arrays. The {dMCLP} {DoA} error is less than $0.5^o$ in most of the cases; i.e., the {DoA} is identified to the nearest integer sampling grid position chosen in the {SRP} computation (for larger enclosures a smaller angular grid can be used). {DoA} computed using the estimated directional component through the iterations of {dMCLP} scheme is found to be better than the estimation using direct microphone signals. Since the source position is computed for each block of the {STFT} signals and the source is assumed stationary, we can combine the source position estimates from successive blocks and then infer a more robust position estimate. This is done by forming a distribution of the estimated DoAs and then identifying the mode of the block-wise {DoA} estimates (non-linear mode filtering). Even the block-wise variation of {DoA} estimates is significantly less in the dMCLP scheme compared to the direct microphone signals based estimation. We have used uniform sampling in the {DoA} angular space for the {SRP}-{PHAT} computation at each node; therefore, the resulting spatial grid resolution decreases with an increase in the distance between the source and microphone array. Accordingly, a small error in {DoA} estimation can cause a larger error in the source position for a larger enclosure and farther sources; this can be seen in the estimation of source position $\bld{s}_3$ using microphone array configuration $\bld{C}_2$. Also, position estimation error depends on the nature of {DoA} errors; a positive error at a node may get compensated by a negative error at a different node. Thus, we can see that the position estimation depends on the placement of nodes, which needs to be optimized for better overall performance. (Also, the placement of furniture/utilities will determine a suitable placement of the ad-hoc array nodes).
\par
\begin{figure}[t]
	\centering
	\begin{minipage}[b]{0.99\linewidth}
		\centerline{\includegraphics[width=3.2in,height=1.05in]{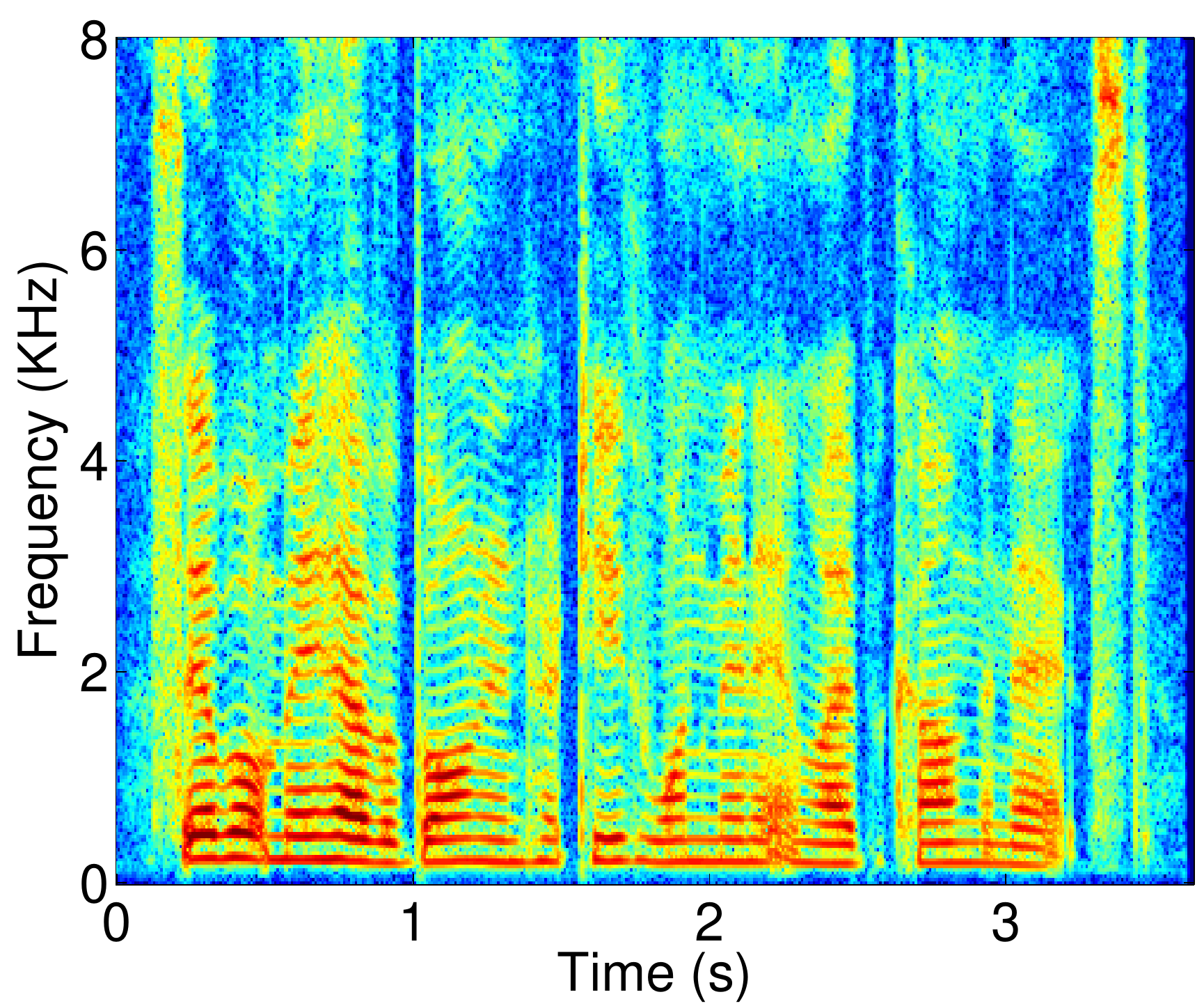}}
		\centerline{(a)}
	\end{minipage}
	\begin{minipage}[b]{0.99\linewidth}
		\centerline{\includegraphics[width=3.2in,height=1.05in]{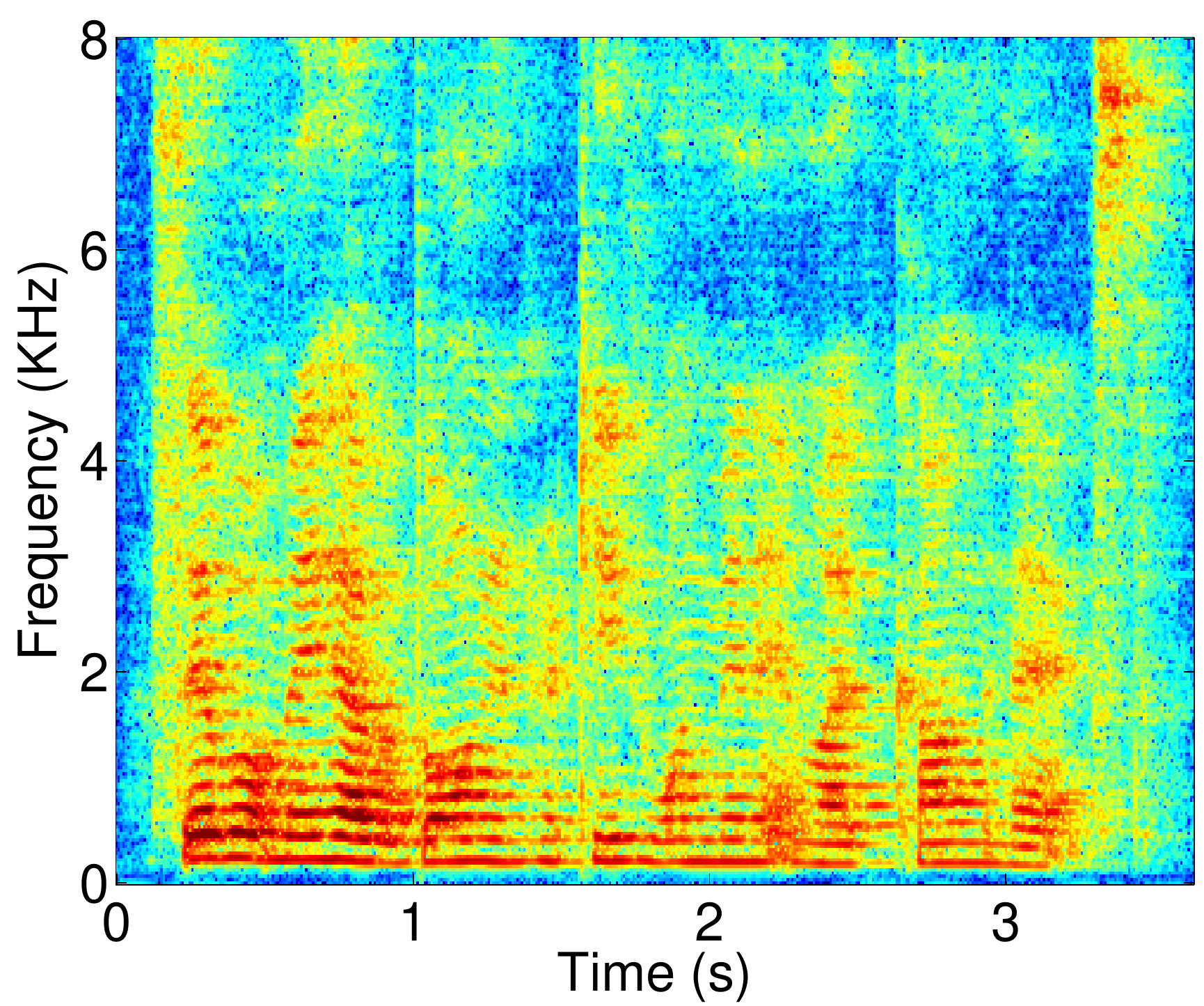}}
		\centerline{(b)}
	\end{minipage}
	\begin{minipage}[b]{0.99\linewidth}
		\centerline{\includegraphics[width=3.2in,height=1.05in]{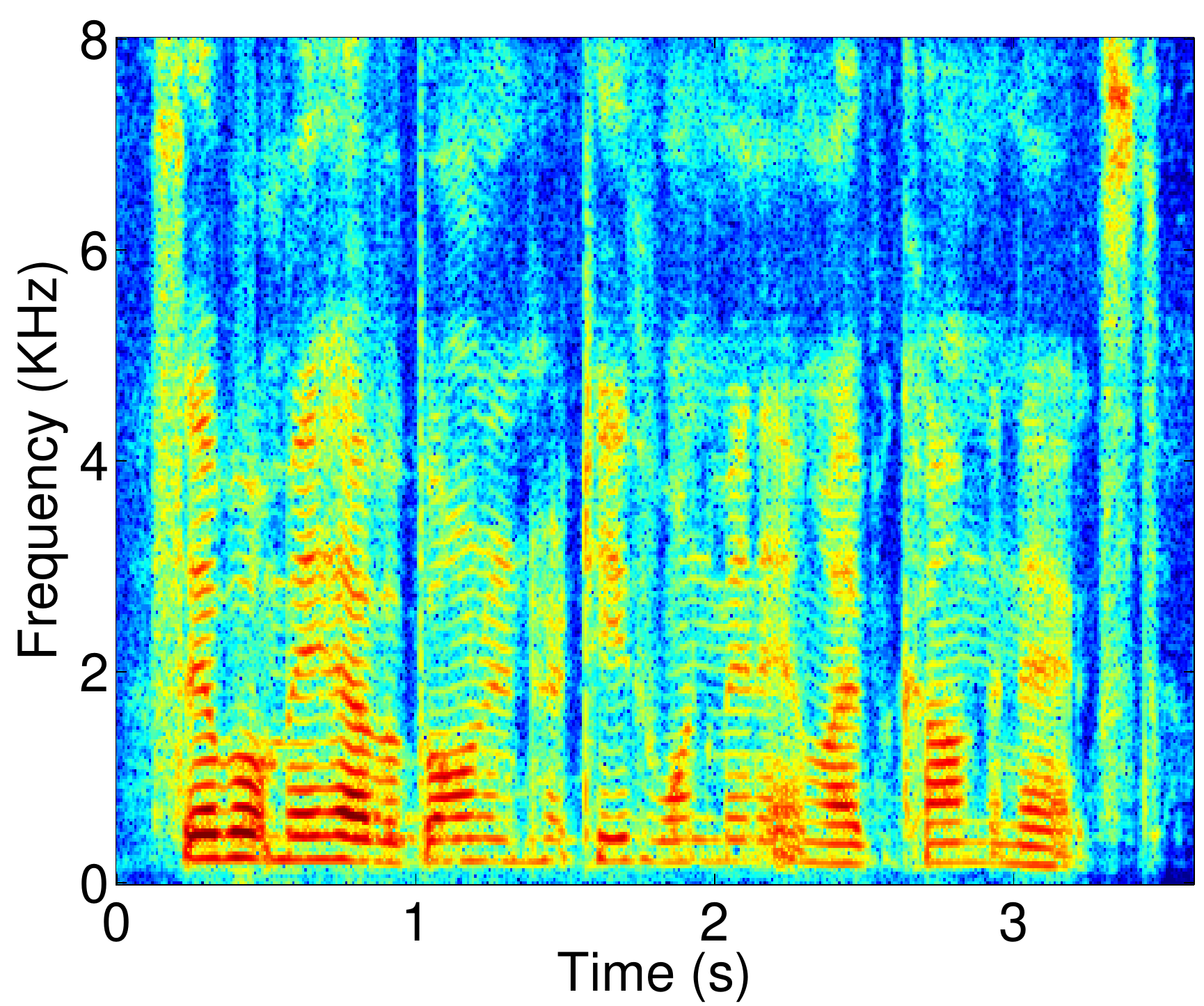}}
		\centerline{(c)}		
	\end{minipage}
	\begin{minipage}[b]{0.99\linewidth}
		\centerline{\includegraphics[width=3.2in,height=1.05in]{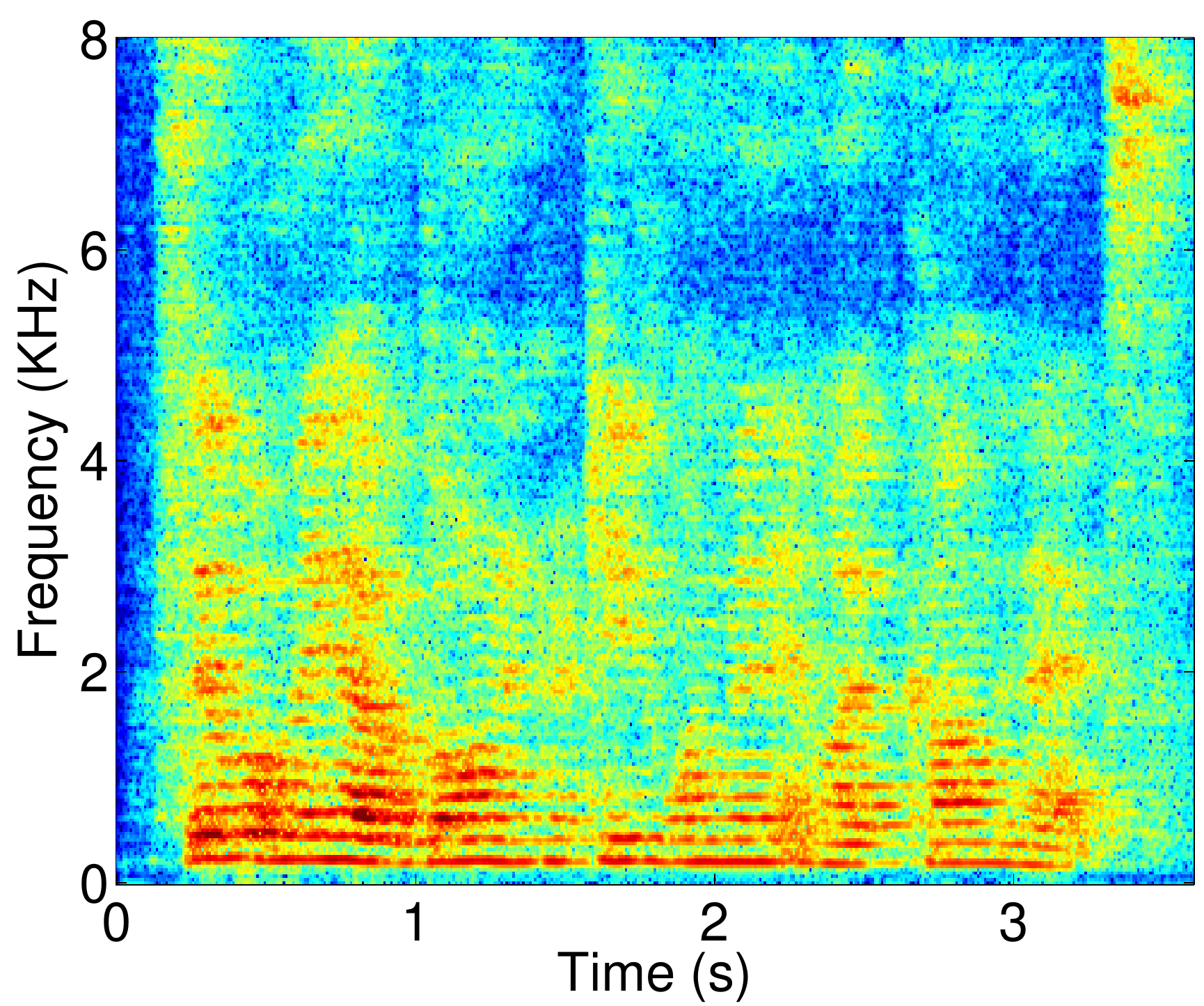}}
		\centerline{(d)}
	\end{minipage}
	\vspace{-5pt}
	\caption{Narrow-band spectrogram of (a) direct component source signal, (b) reverberant mic signal at node $\bld{n}_1$ of $\bld{C}_1$. (c, d) directional and diffuse components estimated at node $\bld{n}_1$. All spectrograms are shown with the same colormap and the large yellow regions in (b,d) indicate the effect of reverberation (diffuse component).}
    \vspace{-5pt}
	\label{fig:specgramIllustration}
\end{figure}

 Spectrogram based illustration of the estimated signals using {dMCLP}, at node $\bld{n}_1$ of configuration $\bld{C}_1$ (nearest node to the source position $\bld{s}_2$), is shown in Fig. \ref{fig:specgramIllustration}. The estimated directional component in \ref{fig:specgramIllustration}(c) is closer to the ground truth source signal shown in \ref{fig:specgramIllustration}(a); the diffuse component shown in \ref{fig:specgramIllustration}(d) comprises of the separated long-term reverb component of the source signal. Sharp onsets are seen only in the estimated directional component. In contrast, diffuse component at the corresponding time frames show temporally smeared spectra. Residual reverberation in the estimated directional component can be observed more at low-frequencies; for example, in the silence region around time instant $1$~s; there is also some distortion at low frequencies as seen in the first harmonic. However, we found that these small distortions are inaudible, through the multi-loudspeaker rendering at the receiver, but audible in the headphone listening.
\par Table \ref{tab:sigEstPerfs} shows the direct component estimation performance at each node quantitatively, based on the average segment-wise {FwSNR} measure \cite{Hu2008Evaluation}.  {FwSNR} is computed for each source position based on the corresponding direct component signal at the microphone. In configuration $\bld{C}_1$, the position $\bld{s}_1$ is approximately equidistant to all the nodes; hence, the FwSNR performance is also nearly uniform at all the nodes. For source positions $\bld{s}_2$ and $\bld{s}_3$, {FwSNR} is found to be best at the nearest node and worst at the farthest node. Similar observations can be made for the microphone array node configuration of $\bld{C}_2$ also. In all the different cases, the dMCLP improvement in {FwSNR} is $\sim 10~$dB at a node nearest to the source compared to the input reverberant microphone signal. The FwSNR improvement indicates the effectiveness of dereverberation, which helps in both source localization and signal intelligibility after reconstruction.
\par The diffuse component estimation is assessed based on the average segmental log spectral distance ({LSD}) measure \cite{Gray1976Distance}, using the reflection component (i.e., input reverb signal minus the original direct component) as the ground truth reference signal \cite{naylor2010speech}. The {LSD} value is $\approx 3~\mbox{dB}$, and it is approximately uniform at all the nodes and also for different source positions in both the nodal configurations. This indicates that the diffuse component estimation is less sensitive to the relative source-microphone placement in the dMCLP scheme.
\begin{table}
	\centering
	\caption{Reverb signal decomposition performance at each node; Frequency weighted SNR (FwSNR) of directional component and log spectral distortion of diffuse component, for different source positions and ad-hoc node configurations.}\label{tab:sigEstPerfs}	
	\begin{tabular}{|c|c|c|c|c|c|}
		\hline
		& & \multicolumn{2}{c|}{Configuration $\bld{C}_1$} & \multicolumn{2}{c|}{Configuration $\bld{C}_2$}\\
		\hline
		\shortstack[c]{Src. \\ Pos.}& Node & \shortstack[c]{Output (Input) \\ FwSNR (dB)} & \shortstack[c]{LSD\\(dB)} &  \shortstack[c]{Output (Input) \\ FwSNR (dB)} & \shortstack[c]{LSD\\(dB)} \\ 
		\hline
		\multirow{4}{*}{\shortstack[c]{$\bld{s}_1$}} 
		& $\bld{n}_1$ & 14.405 (5.253) & 3.459 & \textbf{17.056} (7.366) & 3.170 \\
		& $\bld{n}_2$ & \textbf{15.366} (5.398) & 3.350 & 14.329 (4.695) & 2.582 \\
		& $\bld{n}_3$ & 14.822 (4.887) & 3.466 & 14.739 (5.485) & 2.807 \\
		& $\bld{n}_4$ & 14.798 (4.865) & 3.256 & - & - \\
		\hline
		\multirow{4}{*}{\shortstack[c]{$\bld{s}_2$}} 
		& $\bld{n}_1$ & \textbf{19.500} (8.775) & 3.585 & 14.855 (5.081) & 2.559 \\
		& $\bld{n}_2$ & 16.039 (5.376) & 3.267 & \textbf{17.488} (6.260) & 2.658 \\
		& $\bld{n}_3$ & 12.826 (4.215) & 3.372 & 13.557 (4.301) & 2.820 \\
		& $\bld{n}_4$ & 12.954 (4.686) & 3.335 & - & - \\	
		\hline		
		\multirow{4}{*}{\shortstack[c]{$\bld{s}_3$}} 
		& $\bld{n}_1$ & 11.416 (4.284) & 3.386 & 15.557 (5.482) & 2.531 \\
		& $\bld{n}_2$ & 12.224 (4.045) & 3.314 & 12.586 (4.174) & 2.188 \\
		& $\bld{n}_3$ & \textbf{18.006} (7.102) & 3.600 & \textbf{16.935} (5.329) & 2.150 \\
		& $\bld{n}_4$ & 17.272 (5.903) & 3.337 & - & -\\\hline
	\end{tabular}
	\vspace{-5pt}
\end{table}
\par The effect of radial (angular) grid resolution ($\theta_{step}$) used in the computation of {SRP}-{PHAT}, for the estimation of direct component and also it's {DoA}, is shown in Table \ref{tab:srpGridPerfs}. The results shown are for the source position $\bld{s}_2$ estimation using the configuration $C_1$. As expected, we see that the {dMCLP} scheme does aid in directional component estimation as well as {DoA} estimation. In the present examples, a range of angular grid resolutions from $1^o$ to $5^o$ has minimal effect on the position estimation accuracy and the directional component signal estimation. However, increasing the grid resolution to $10^o$ shows a significant increase in the source position error; the {FwSNR} also decreases by up to $1.5~dB$. We note that the inter microphone spacing at each array node is more important for the accuracy of {DoA} and hence the source position estimation.
\begin{table}[t]
	\centering
	\caption{Effect of angular grid resolution of SRP-PHAT on the estimation performance for configuration $C_1$.}\label{tab:srpGridPerfs}	
\begin{tabular}{|c|c|c|c|c|c|}
	\hline
	& Node & \shortstack[c]{$1^o$} &  \shortstack[c]{$2^o$ } & \shortstack[c]{ $5^o$} & \shortstack[c]{$10^o$}\\ 
	\hline
	\multirow{4}{*}{\shortstack[c]{FwSNR \\ o/p}} 
	& $\bld{n}_1$ & 19.500 & 19.692 & 19.500 & 18.063\\
	& $\bld{n}_2$ & 16.039 & 16.071 & 15.926 & 15.926\\
	& $\bld{n}_3$ & 12.826 & 12.829 & 12.658 & 12.166\\
	& $\bld{n}_4$ & 12.954 & 12.954 & 12.788 & 12.788\\\cline{3-6}
	\hline
	\multirow{4}{*}{\shortstack[c]{Angle \\ error}} 
	& $\bld{n}_1$ & 0.53$^o$ & 0.46$^o$ & 0.54$^o$ & 4.46$^o$\\
	& $\bld{n}_2$ & 0.31$^o$ & 0.51$^o$ & 2.51$^o$ & 2.51$^o$\\
	& $\bld{n}_3$ & 1.53$^o$ & 0.53$^o$ & 0.46$^o$ & 4.53$^o$\\
	& $\bld{n}_4$ & 0.20$^o$ & 0.20$^o$ & 2.20$^o$ & 2.20$^o$ \\\cline{3-6}\hline
	\multicolumn{2}{|c|}{RMSPE (cm)} & 5.83 cm & 2.17 cm & 3.08 cm & 15.03 cm\\\hline
\end{tabular}	
\vspace{-5pt}
\end{table}
	\subsection{Rx Spatial Reconstruction}
We examine two schemes of spatial rendering at the receiver, using either (i) loudspeakers, or (ii) head-phone. In each of these modalities there are certain advantages and also challenges. Since our goal is to meet a listener's perceptual impression of ``spatial speech'', we consider measuring the subjective quality using a MUSHRA like scoring scheme. The listeners score (i) naturalness and intelligibility of speech, and (ii) perceived source position. 
\subsubsection{Loud-speaker setup}
We consider a four-channel loudspeaker (LS) setup, in which the speakers are placed at $\{45^o,~135^o,~225^o,~\mbox{and}~315^o\}$ of the listener front position and approximately at equidistant from the actual listener position. (Any deviation from this symmetric configuration can be compensated in the signal rendering method). The loudspeakers are powered by a custom built power amplifier. The input signals to the power amplifier are fed from ``Focusrite Scarlet 18i20'' multichannel D-A converter. The gains of the loudspeakers are adjusted (using a sound level meter and a white nose source as the calibration signal) to be within $\pm 0.2~$dB range about the mean listening level of $65~$dB at the listener position. 
We consider a fixed VL position $\bs{\ell}$ in the Tx enclosure, as shown in Fig. \ref{fig:recordingSetup}, having $(x,y)$ coordinates of $(4,2.85)$~m. The (direction, distance) parameters of the three sources for the chosen VL position, taking the direction of listener-$\bs{\ell}$ right ear as $0^o$, are $(90^o,1.0\mbox{m}), ~(160^o,1.12\mbox{m}), ~\mbox{and} ~(62.4^o,2.48\mbox{m})$ respectively.
\subsubsection{Signal Reconstruction (rendering)} \label{sec:sigrepssc}
We consider two experiments to study the perceived quality of spatial reconstruction. In Expt-1, we study only the direct component reconstruction without the additional diffuse component; in Expt-2, the complete reverberant signal estimated at the VL position $\bs{\ell}$ is reconstructed at the receiver. These two listening tests are chosen to explore the importance of diffuse signal component reconstruction at the receiver. 

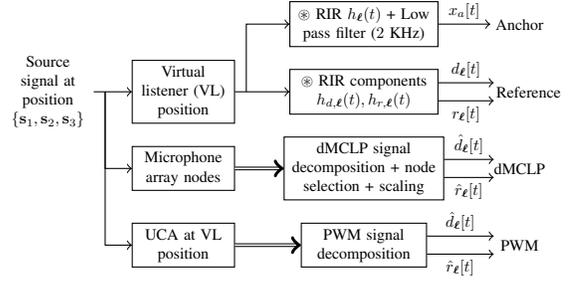
\begin{figure}[t]
	\centering
	\begin{tikzpicture}[scale=.6,every node/.style={scale=.6}]
	\node[text width=0.7in,align=center] (SS) at (-7.5,0) {Source signal at position $\inCurly{ \bld{s}_1,\bld{s}_2,\bld{s}_3 }$};
	\node[draw,text width=0.8in,align=center] (VLRIR) at (-4.5,0) {Virtual listener (VL) position };
	\draw[->] (SS) -- (VLRIR);
	\node[draw,text width=1.2in,align=center] (SD) at (-0.5,0) {$\circledast$ RIR components $h_{d,\bs{\ell}}(t),h_{r,\bs{\ell}}(t)$ };
	\draw[->] (VLRIR) -- (SD);		
	\node[draw,text width=1.2in,align=center] (LPF) at (-0.5,1.5) {$\circledast$ RIR $h_{\bs{\ell}}(t)$ + Low pass filter ($2$~KHz)};
	\draw[->] ([xshift=0.1in]VLRIR.east) |-  (LPF);
	\node[] (dl) at (2.5,0.2) {}; 
	\node[] (rl) at (2.5,-0.2) {}; 
	\draw[->] (SD.east |- dl) --node[above]{$d_{\bs{\ell}}[t]$} (dl);
	\draw[->] (SD.east |- rl) --node[below]{$r_{\bs{\ell}}[t]$} ( rl);
	\node[] at (3.15,0) {Reference};		
	\node[] (A) at (2.95,1.5) {Anchor};
	\draw[->] (LPF) --node[above]{ $x_a[t]$} (A);
	
	\node[draw,text width=0.8in,align=center] (NRIR) at (-4.5,-1.7) {Microphone array nodes};
	\draw[->] ([xshift=0.1in]SS.east) |- (NRIR);
	\node[draw,text width=1.3in,align=center] (DSD) at (-0.5,-1.7) {dMCLP signal decomposition + node selection + scaling};
	\draw[->,double] (NRIR) -- (DSD);
	\node[] (dn) at (2.5,-1.5) {}; 
	\node[] (rn) at (2.5,-1.9) {}; 
	\draw[->] (DSD.east |- dn) --node[above]{${\hat d}_{\bs{\ell}}[t]$} (dn);
	\draw[->] (DSD.east |- rn) --node[below]{${\hat r}_{\bs{\ell}}[t]$} (rn);				
	\node[] (D) at (2.95,-1.7) {dMCLP};
	
	\node[draw,text width=0.8in,align=center] (UCARIR) at (-4.5,-3.4) {UCA at VL position};
	\draw[->] ([xshift=0.1in]SS.east) |- (UCARIR);
	\node[draw,text width=1in,align=center] (PSD) at (-0.5,-3.4) {PWM signal decomposition};
	\draw[->,double] (UCARIR) -- (PSD);
	\node[] (dn2) at (2.5,-3.2) {}; 
	\node[] (rn2) at (2.5,-3.6) {}; 
	\draw[->] (PSD.east |- dn2) --node[above]{${\hat d}_{\bs{\ell}}[t]$} (dn2);
	\draw[->] (PSD.east |- rn2) --node[below]{${\hat r}_{\bs{\ell}}[t]$} (rn2);				
	\node[] (P) at (2.95,-3.4) {PWM};		
	\end{tikzpicture}
	\vspace{-5pt}
	\caption{Stimuli generation for the MUSHRA based evaluation.}\label{fig:mushraSignals} 	
	\vspace{-5pt}
\end{figure}

The scheme used to generate the listening test stimuli is shown in Fig. \ref{fig:mushraSignals}. For the ground truth, the direct component signal $d_{\bs{\ell}}[t]$, and the diffuse late reverb signal $r_{\bs{\ell}}[t]$ at the VL position $\bs{\ell}$ are computed using the known RIR at the transmitter enclosure. These are then used to generate the oracle reference signals in perceptual scoring. The dMCLP approach is also compared with the PWM based method for direct, diffuse component estimation as in \cite{thiergart2014informed}. In the PWM method, we consider signal acquisition using a separate $4$-channel UCA (radius=$0.1~m$) placed at the VL position. PWM method is implemented using the known ground truth speech activity (for noise PSD estimation) and the known source DoA. The ground truth reverberant signal at the VL position $x_{\bs{\ell}}[t]$ is also low pass filtered to $2~$KHz and used to generate the anchor signal $x_a[t]$. In Expt-1, only the direct component signal $d_{\bs{\ell}}[t]$ is rendered from the given source direction, and in Expt-2, the complete reverberant signal is rendered using both the components $d_{\bs{\ell}}[t]$ and $r_{\bs{\ell}}[t]$. In Expt-1, the anchor signal $x_a[t]$ is rendered from the ground truth source direction, whereas in Expt-2, $x_a[t]$ is synthesized as a diffuse signal (no directional component) using the method discussed in sec. \ref{sec:diffuseRender}. 
\begin{figure}[t]
	\centering
	\includegraphics[width=2.5in,height=1.6in]{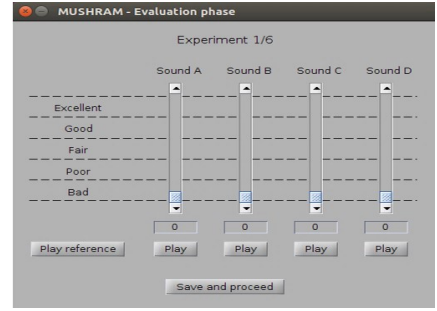}
	\vspace{-5pt}
	\caption{MATLAB GUI used for MUSHRA test of blind comparison and scoring the quality/direction of the test signals in comparison to the reference at the bottom left.}\label{fig:mushraGui}
	\vspace{-5pt}
\end{figure}
\par The graphical user interface (GUI) for the MUSHRA like perceptual evaluation is shown in Fig. \ref{fig:mushraGui}. Two test signals, a hidden reference signal and an anchor signal are assessed for their quality on a continuous scale of $0-100$, in comparison to the explicit reference signal. Uniformly spaced qualitative markers {Excellent, Good, Fair, Poor, Bad} are provided for the listener. The listeners are asked to rate the four signals in terms of signal quality, intelligibility, source direction, and spatialization (overall spatial impression) with respect to the reference signal. The listeners are advised to play the reference signal and the test signals as many times as they desire before fixing the evaluation scores. Five listeners, students in the age group of $20-35$ years, participated in the listening experiment. Two speech signals, one each from a male and a female speaker, taken from the TIMIT database are used as the source signals. We consider the three source positions shown in Fig. \ref{fig:recordingSetup}(a) for the evaluation, corresponding to a total of six sets of stimuli ($3$ source positions $\times$ $2$ signals (male/female)). Each listener repeats the two experiments, Expt-1, Expt-2, for each of the six sets of stimuli.
\begin{figure}[t]
	\begin{minipage}[b]{0.49\linewidth}
		\centerline{\includegraphics[width=1.5in,height=1.35in]{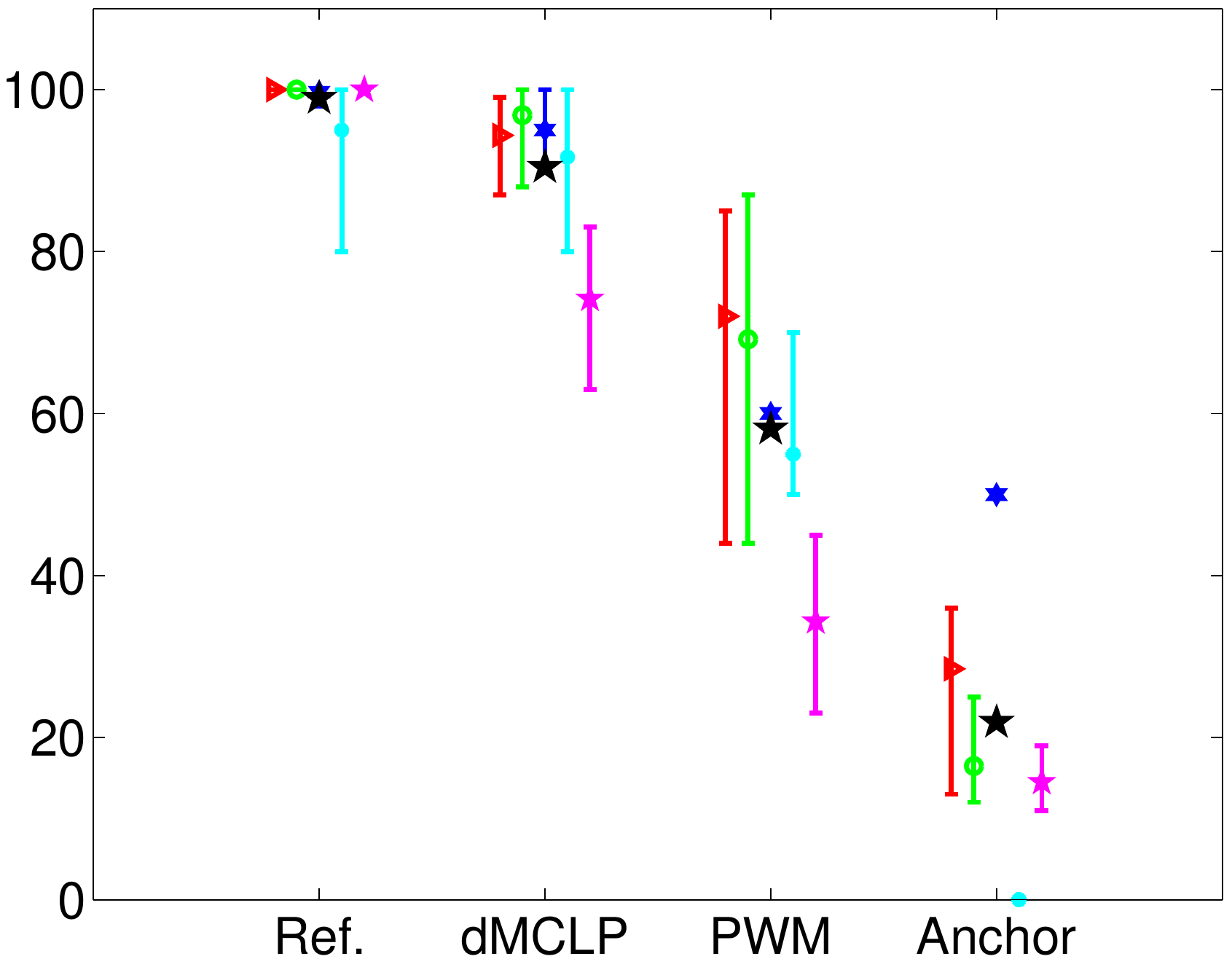}}
		\centerline{(a)}
	\end{minipage}
	\begin{minipage}[b]{0.49\linewidth}
		\centerline{\includegraphics[width=1.5in,height=1.35in]{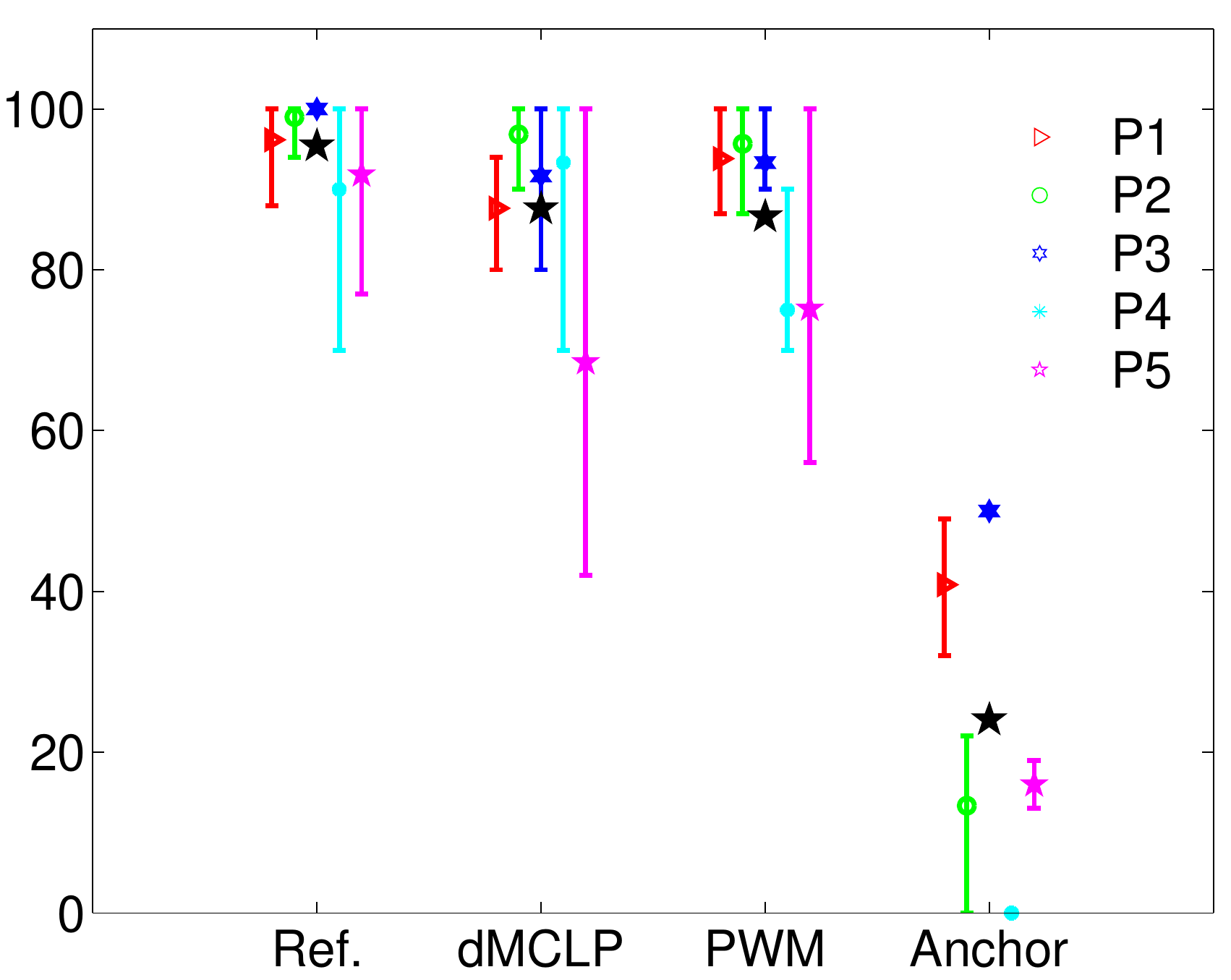}}
		\centerline{(b)}
	\end{minipage}
	\vspace{-5pt}
	\caption{LS rendering: MUSHRA scores for (a) Expt-1 and (b) Expt-2. The lower, upper bars and the symbol at the center show the min, max and mean values respectively. Star symbol in black color shows the average MUSHRA score across all listeners and the test stimuli.}\label{fig:dirCompMEval}
	\vspace{-5pt}
\end{figure}
\par Listeners are asked to rate the quality of the randomized test stimuli, give a verbal description of the perceptual impression of each of the stimuli, and also describe the spatial impression. The verbal description was consistent across different listeners. In Expt-1, two stimuli (hidden reference signal and dMCLP estimated signal) are similar to each other and closer to the explicit reference; one stimulus (PWM method) has higher residual reverberation, and the fourth stimulus is distinct from the other stimuli with a poor quality as expected (anchor signal). In Expt-2, two of the stimuli (hidden reference signal and dMCLP estimated signal) are similar to each other and closer to the explicit reference; the third stimulus (of the PWM method) is judged ``closer'' to the listener or ``louder'' compared to the explicit reference and better in terms of quality; however the anchor stimulus is distinct with a poorer quality. The perception of source direction is also stable and consistent across all stimuli.
\par Fig. \ref{fig:dirCompMEval} shows the MUSHRA scores for the five listeners, individually. Fig. \ref{fig:dirCompMEval}(a) shows the scores of Expt-1. We see that the hidden reference signal is identified correctly by four listeners, but one listener (shown in cyan color) confused it with the dMCLP estimated signal. The anchor signal is identified correctly by all listeners and rated to be ``poor/bad''. The proposed dMCLP based reconstruction is ranked closer to the reference, consistently for all the six conditions, and by all the five listeners. The PWM approach is ranked worse than dMCLP and assessed closer to the anchor signal. The qualitative description also varied from ``poor'' to ``good'', due to the excess residual reverberation in the estimated direct component signal of the PWM approach.
\par Fig. \ref{fig:dirCompMEval}(b) shows the MUSHRA scores for Expt-2. Anchor signal is consistently rated low by all listeners, though the qualitative description differed across the subjects. The hidden reference signal is not identified correctly by all the listeners for all the stimuli, which can be noted from the spread in listener ratings for the hidden reference signal. MUSHRA evaluation requires at least one of the test signals to be scored $100$. The plot indicates that the hidden reference is confused with the test signals in some cases, indicating the high quality of reconstructed signals. We see that the dMCLP and PWM approaches are comparable, and the subjective rating differences are marginal across the listeners (PWM utilizes known ground truth, like an oracle). Direct component estimated in PWM consists of excess residual reverberation, which gives rise to increased loudness. Rendering the diffuse component also creates an impression of the source being closer and better intelligible than the reference signal. This caused confusion in some cases, and the listeners gave a score of $100$ to the PWM stimuli instead of the hidden reference signal. This effect is confirmed through the verbal description of the subjects also.
\par We note that the spatial location of the reconstructed signal is computed using two-channel microphone arrays first, and the signals are used for the listener position in the dMCLP scheme. In contrast, for the PWM approach, we use a four-channel microphone array placed at the ground truth VL position and assume the ground truth source direction is known a-priori. Despite this computational advantage of the PWM method, the spatial reverberant signal reconstruction is comparable between the two approaches; however the direct component only reconstruction is better using the dMCLP approach. Mapping the signals estimated using the PWM approach to a different VL position would result in an inferior spatial image due to the excess residual reverberation in the directional component.
\subsubsection{Effect of delay parameter}\label{sec:effect_delay}
The analysis parameter $D$ at the transmitter is an important parameter determining the quantum of separation between the direct and diffuse components. These two separated signals are reconstructed differently at the receiver to create the natural listening effect to the Rx listener. Hence, we need to study the effect of the analysis delay parameter $D$ on the perceived quality of spatial reconstruction of the reverberant signal. We consider the source at position $\bld{s}_2$ in Fig. \ref{fig:recordingSetup} for the evaluation. Two test signals, one each from a male and a female speaker, are used to generate the stimuli signals. Thus, each listener scores two sets of test stimuli, and in each set, listens to a total of six test stimuli: hidden reference signal, anchor signal, and four signals corresponding to the dMCLP analysis parameters of $D=\{1,\dots,4\}$. The directional and diffuse component signals estimated at the node $\bld{n}_2$, which is nearest to the source position $\bld{s}_2$, are used for spatial reconstruction to determine the optimum $D$.
\par
\begin{figure}[t]
	\centering
	\begin{minipage}[b]{0.48\linewidth}
		\centerline{\includegraphics[width=1.5in,height=1.35in]{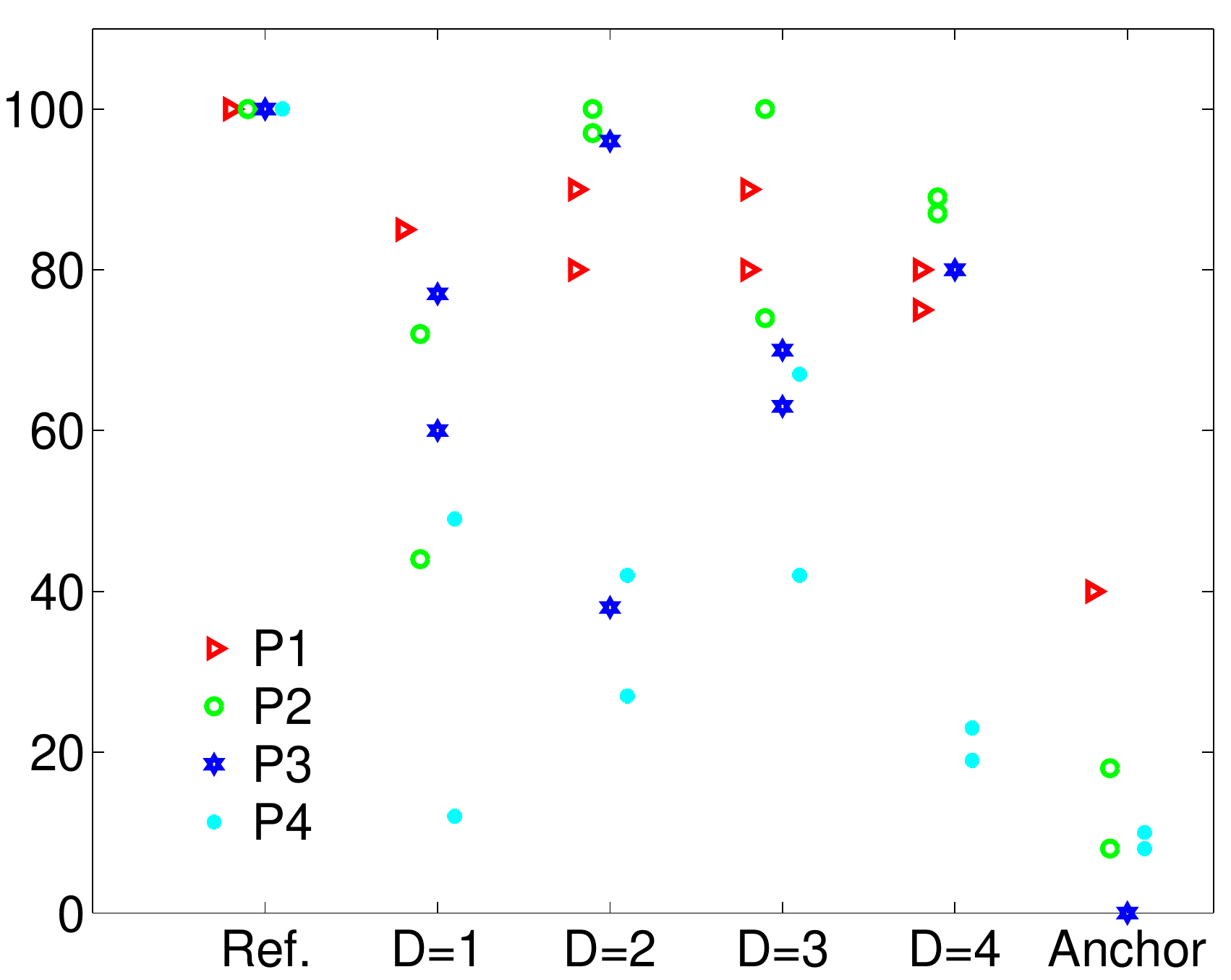}}
		\centerline{(a)}
	\end{minipage}
	\begin{minipage}[b]{0.48\linewidth}
		\centerline{\includegraphics[width=1.5in,height=1.35in]{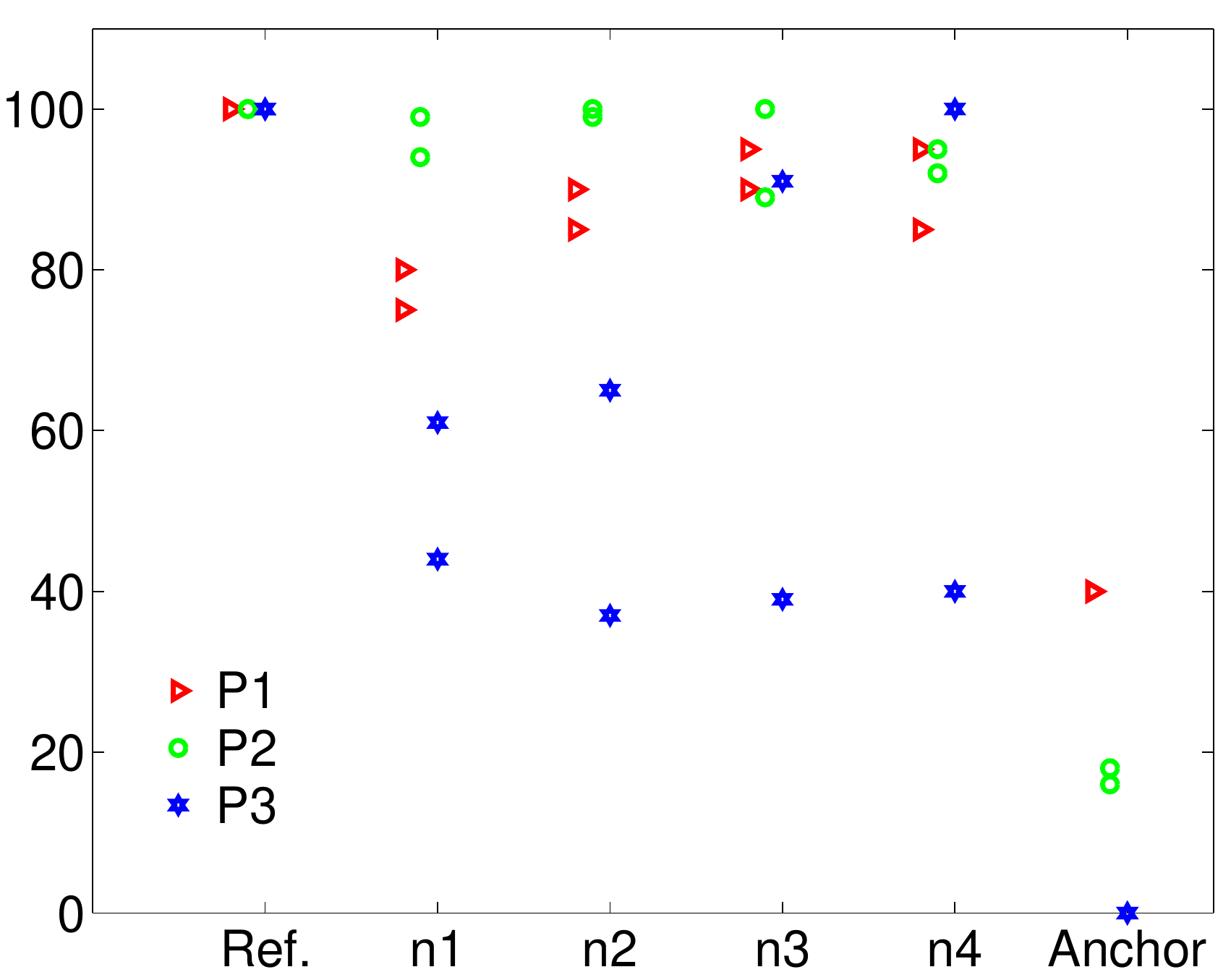}}
		\centerline{(b)}
	\end{minipage}	
	\vspace{-5pt}
	\caption{LS rendering: (a) MUSHRA scores for the MCLP analysis parameter $D$. (b) MUSHRA scores for the selected node $n^*$ at the transmitter.}\label{fig:DNperc}
	\vspace{-10pt}
\end{figure}

 The listening experiment results are shown in Fig. \ref{fig:DNperc}(a). MUSHRA scores from one listener, who identified the hidden reference wrongly, are discarded, and the scores for the remaining four listeners are shown in the plot. The quality scores are better for delay parameter $D=2$ compared to higher $D$, and $D=1$ has a poor perceived quality. As we have noted in the Tx signal estimation experiments, the choice of $D=1$ results in distortion of the directional signal component itself, which leads to the degradation of perceived quality in spatial reconstruction. On the other hand, $D=2$ case has better signal estimation, resulting in a perceived spatial impression closer to the reference signal. A higher value of $D$ includes more residual reverberation in the estimated directional component, which is found to result in a different spatial image than the explicit reference and poorer perceived quality along with timbral artifacts. 
\subsubsection{Effect of node selection}
For spatial reconstruction, we hypothesized that the signals at the array node nearest to the source position would be the best because of its high SRR. However, while reconstructing at the receiver, the perceptual quality and spatial impression become more important than just the SRR measure. Hence, we examine the selection of different nodes perceptually also. Similar to the experiment earlier, we consider the source at position $\bld{s}_2$. Each listener listens to two sets of stimuli, with a total of six signals in each set: hidden reference signal, anchor signal, and four test signals corresponding to the direct and diffuse components estimated at the nodes $\{\bld{n}_1,\dots,\bld{n}_4\}$ and the parameter $D=2$.
\par Fig. \ref{fig:DNperc}(b) shows the effect of the selected array node for spatial reconstruction. The MUSHRA test samples based on the four nodes are randomized and presented to the five listeners individually. We omitted the scores from two listeners who wrongly identified the hidden reference, and show the scores for only three listeners (the omitted listeners probably needed more concentration or more example stimuli). We see that there is a preference to nodes $\bld{n}_2$ and $\bld{n}_3$, where node $\bld{n}_2$ is nearest to the source position $\bld{s}_2$ compared to the other nodes. This is expected since the signal estimation is better at the nodes nearer to the source position compared to farther nodes, where the residual reverberation can be high. Also, the strength of the direct component at a farther node is poorer compared to the reflections (low SRR). Because of this, the scaling of the direct component, given in eqn. \eqref{eqn:signaltransformation}, to approximate the signal at the VL position $\bs{\ell}$, results in amplification of the residual reverberation component, resulting in timbral artifacts. Hence the spatial reconstruction for a VL position based on the nearest array node does provide the best perceptual reconstruction also.
\subsubsection{Binaural Reconstruction}
We also studied the binaural rendering of spatial speech signals. We adopt the spatial reconstruction scheme proposed in \cite{keuch2008directional} for binaural rendering. The rendering scheme is shown in Fig. \ref{fig:binrend}. In this scheme, the playback signals are first generated for LS rendering and then convolved with the HRIRs corresponding to the LS directions and added together to generate the two-channel binaural input signals. The HRIR essentially introduces relative ITD and ILD cues between the left and right ears, which the LS signals would have induced for a listener at the center of the LS configuration. The chosen angles for the HRIR is to suit the earlier LS position simulation, so that we can use the same VBAP calculation for the rendering. (A different configuration of loudspeakers will require modified VBAP calculation and subsequent change in HRIRs for binaural rendering). We repeat the perceptual evaluation of signal reconstruction experiments as in sec. \ref{sec:sigrepssc} using the binaural rendering scheme also.
\begin{figure}[t]
	\centering
	\begin{tikzpicture}[scale=0.55,every node/.style={scale=0.55}]
	
	\node[] (d) at (-4.5,2.5) {$d^*[t]$};
	\node[] (r) at (-4.5,0.5) {$r^*[t]$};		
	
	\node[draw,minimum height=0.4in,text width=0.9in,align=center] (v) at (-2.5,2.5) {VBAP $\inSqBrackets{g_1^*,\dots,g_S^*}$};
	\node[draw,minimum height=0.4in,text width=0.9in,align=center] (D) at (-2.5,0.3) {Decorrelation $c_s[t]/\sqrt{S},~\forall~s$};
	
	\node[draw,circle] (s1) at (0.2,4) {\tiny $\Sigma$};
	\node[draw,circle] (s2) at (-0.2,-1) {\tiny $\Sigma$};
	\node[draw,circle] (s3) at (1.3,0.7) {\tiny $\Sigma$};
	\node[draw,circle] (s4) at (1.3,2.3) {\tiny $\Sigma$};						
	
	\draw[->] ([yshift=-3mm]D.east) -- ([yshift=-3mm,xshift=2mm]D.east) |- (s2);
	\draw[->] ([yshift=-1mm]D.east)  -| (s3.south);
	\draw[->] ([yshift=1mm]D.east) -- ([yshift=1mm,xshift=15mm]D.east) -- ([yshift=10mm,xshift=15mm]D.east) -| (s4.south);
	\draw[->] ([yshift=3mm]D.east) -| (s1.south);
	
	\draw[->] ([yshift=-3mm]v.east) -| (s2.north);
	\draw[->] ([yshift=-1mm]v.east) --([yshift=-1mm,xshift=9mm]v.east) |- (s3.west);
	\draw[->] ([yshift=1mm]v.east) --([yshift=1mm,xshift=15mm]v.east) |- (s4.west);
	\draw[->] ([yshift=3mm]v.east) --([yshift=3mm,xshift=2mm]v.east) |- (s1.west);
	
	\draw[->,blue] (d) -- (v);
	\draw[->,red] (r) -- (D);
	
	\node[draw,text width=0.9in,align=center] (h1) at (3.5,4) {HRIR($45^o$) $\circledast h_L[t],h_R[t]$};
	\node[draw,text width=0.9in,align=center] (h2) at (3.5,-1) {HRIR($315^o$) $\circledast h_L[t],h_R[t]$};
	\node[draw,text width=0.9in,align=center] (h3) at (3.5,0.7) {HRIR($225^o$) $\circledast h_L[t],h_R[t]$};
	\node[draw,text width=0.9in,align=center] (h4) at (3.5,2.3) {HRIR($135^o$) $\circledast h_L[t],h_R[t]$};
	
	\draw[->] (s1) -- (h1);
	\draw[->] (s2) -- (h2);
	\draw[->] (s3) -- (h3);
	\draw[->] (s4) -- (h4);	
	
	\node[draw,circle] (b1) at (6.5,2.5) {$+$};			
	\node[draw,circle] (b2) at (6.5,0.5) {$+$};
	\draw[->,red] ([yshift=0.2cm]h1.east) -| (b1);
	\draw[->,red] ([yshift=0.2cm]h4.east) -- (b1);		
	\draw[->,red] ([yshift=0.2cm]h3.east) -- ([yshift=0.2cm,xshift=0.5cm]h3.east) -- ([yshift=0.4cm,xshift=0.5cm]h3.east) -- (b1);		
	\draw[->,red] ([yshift=0.2cm]h2.east) -- ([yshift=0.2cm,xshift=0.8cm]h2.east) -- ([yshift=0.7cm,xshift=0.8cm]h2.east) -- (b1);
	
	\draw[->,blue] ([yshift=-0.2cm]h1.east) -- ([yshift=-0.2cm,xshift=0.8cm]h1.east) -- ([yshift=-0.7cm,xshift=0.8cm]h1.east) -- (b2);
	\draw[->,blue] ([yshift=-0.2cm]h4.east) --([yshift=-0.2cm,xshift=0.5cm]h4.east) -- ([yshift=-0.4cm,xshift=0.5cm]h4.east) -- (b2);
	\draw[->,blue] ([yshift=-0.2cm]h3.east) -- (b2);
	\draw[->,blue] ([yshift=-0.2cm]h2.east) -| (b2);
	
	\node[] (o1) at (8,2.5) {$s_L[t]$};
	\node[] (o2) at (8,0.5) {$s_R[t]$};			
	\draw[->,red] (b1) -- (o1);\draw[->,blue] (b2) -- (o2);
	\end{tikzpicture}
	\vspace{-5pt}
	\caption{Binaural rendering using loudspeaker signals convolved with HRIRs of the four LS directions and summed to generate the binaural signals.} \label{fig:binrend}	
\end{figure}
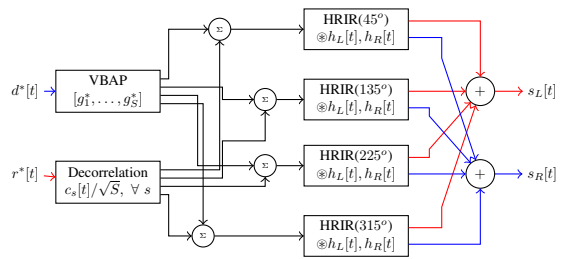
\begin{figure}[t]
	\begin{minipage}[b]{0.49\linewidth}
		\centerline{\includegraphics[width=1.5in,height=1.35in]{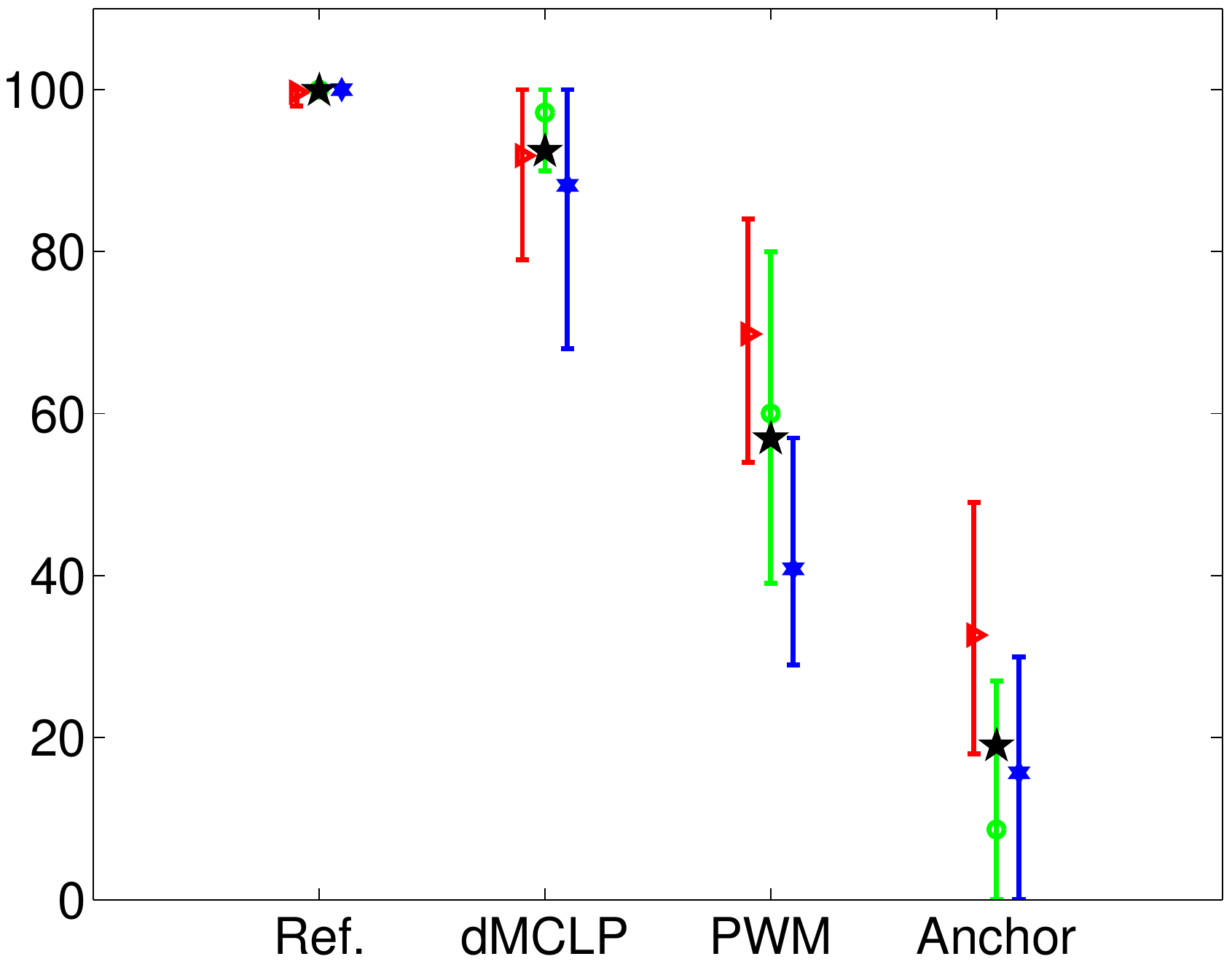}}
		\centerline{(a)}
	\end{minipage}
	\begin{minipage}[b]{0.49\linewidth}
		\centerline{\includegraphics[width=1.5in,height=1.35in]{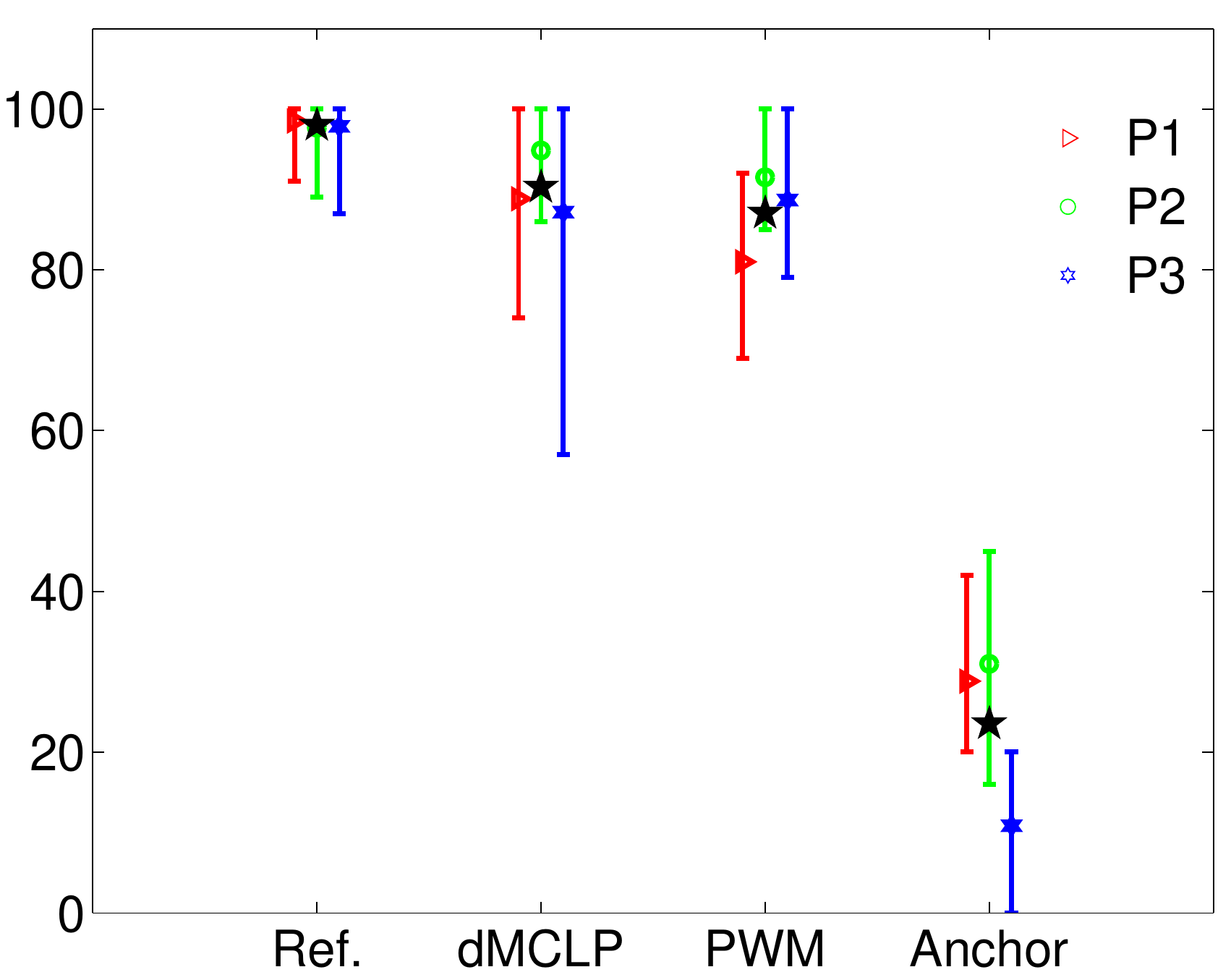}}
		\centerline{(b)}
	\end{minipage}
	\vspace{-5pt}
	\caption{Binaural rendering: MUSHRA scores for (a) Expt-1 and (b) Expt-2. The colors show results for the three listeners, and the black $\bigstar$ shows the average score across the listeners and test files. The lower bar, upper bar and the symbol at the center show the min, max and mean values respectively.
	}\label{fig:dirCompMEvalBin}
	\vspace{-5pt}
\end{figure}
\par Fig. \ref{fig:dirCompMEvalBin} shows the MUSHRA scores for the two experiments, corresponding to three listeners. The MUSHRA scores tend to be similar to the LS rendering case. In the direct component reconstruction evaluation of Expt-1, the hidden reference and the directional anchor are identified correctly by all the listeners; the dMCLP based reconstruction is rated better than the PWM approach. The spread in the dMCLP scores is higher for binaural reconstruction compared to the earlier LS rendering case (shown in Fig. \ref{fig:dirCompMEval}(a)). This may be because the slight distortions due to dMCLP estimation may become audible when the signals are presented over the headphones and inaudible through the direct LS presentation. In the reverberant signal reconstruction evaluation of Expt-2, the hidden reference signal is often confused with the dMCLP or PWM stimuli, again similar to the LS reconstruction case (Fig. \ref{fig:dirCompMEval}(b)); the dMCLP, PWM oracle test stimuli are rated to be closer to each other and also to the reference signal\footnote{Another alternative to binaural rendering is to directly convolve the directional component with the left/right HRIRs and then add the decorrelated diffuse component to both the left and right signals. However, externalization of the source in binaural perception becomes important in this.}.
\section{Virtual Spatial Scene Rendering}
\begin{figure}[t]
	\centering
	\begin{tikzpicture}
	\node[] at (0,0) {\includegraphics[width=3in,height=1.6in]{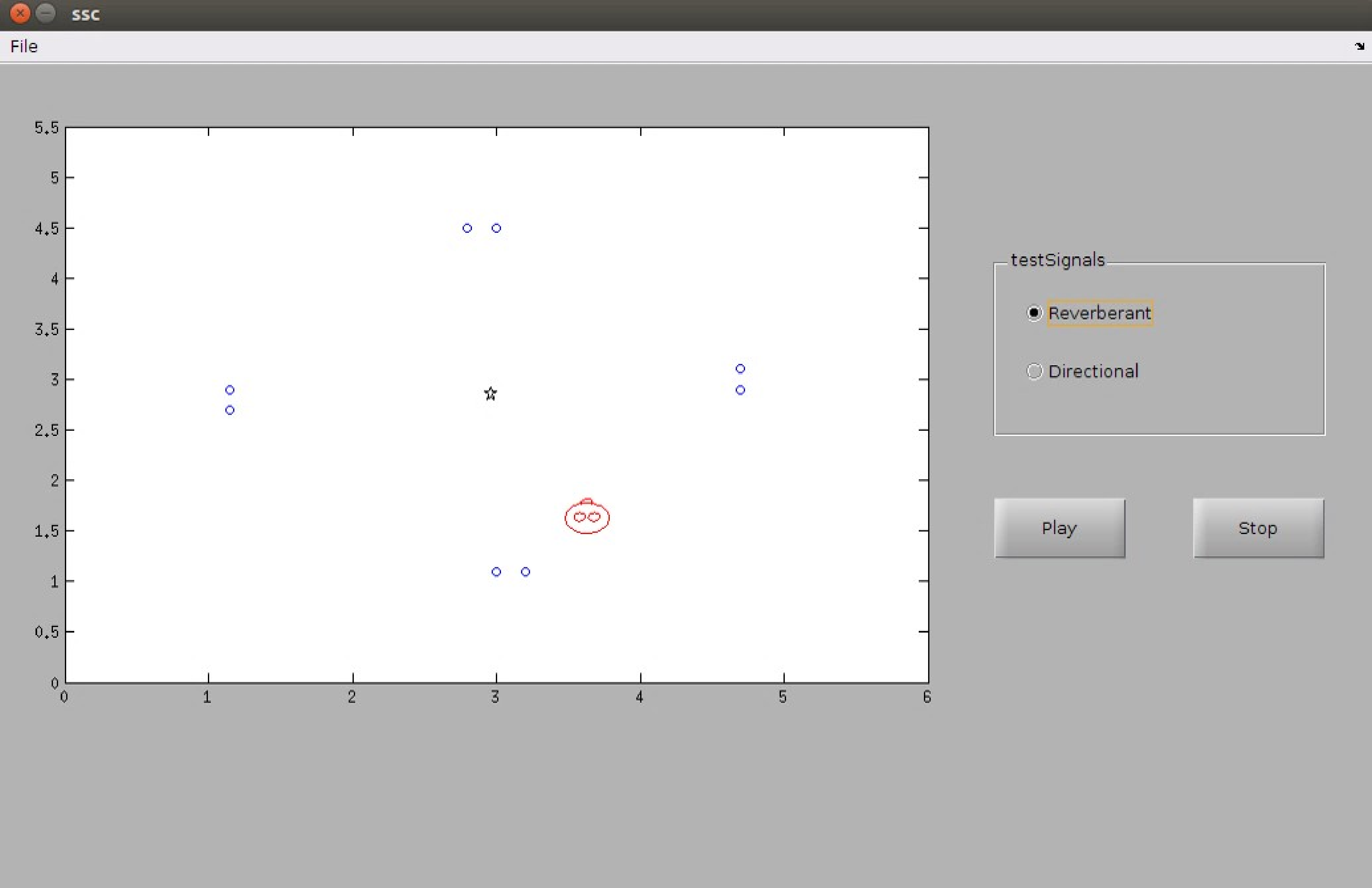}};
	\node[] at (-0.6,-0.6) {\tiny $\bld{s}$};
	\node[] at (-1.25,0.25) {\tiny $\bs{\ell}$};
	\end{tikzpicture}	
	\vspace{-5pt}
	\caption{GUI for spatial scene rendering test. Source position $\bld{s}$ is shown in color 'red', and the virtual listener position $\bs{\ell}$ is shown in 'black'. The listener is free to choose virtual listener position using a mouse click.}\label{fig:ssc_ss}
	\vspace{-5pt}
\end{figure}
We conducted informal listening experiments of transmitter space reproduction for the spatial speech communication, using a single stationary source case. Fig. \ref{fig:ssc_ss} shows the interactive GUI using MATLAB for this listening test. Source position and the directional, diffuse components at the best array node are estimated using the scheme of sec. \ref{sec:spatialAnalysis}. On the GUI, the estimated source position and the microphone array node positions are shown such that the listener at the receiver is aware and may choose to alter the VL position $\bs{\ell}$ in the transmitter space. The initial VL position is chosen to be close to the center of the transmit enclosure and oriented towards north. The real listener at Rx can change the VL position by using the mouse control. The listener is also allowed to choose between spatial reconstruction using (i) directional component only (as in Expt-1) or (ii) the reverberant signal, which has both directional and diffuse components (as in Expt-2)\footnote{We can provide as ground-truth an explicit reference signal recorded at the VL position $\bs{\ell}$ to compare the overall acoustic scene perception and to get an overall MUSHRA score}. For the initial VL position, the source is behind the virtual listener; the same is perceived by the real listener. Changing the VL position radially does provide a change in the perceived source position, consistently, for both direct component and reverberant signal reconstruction. In the case of reverberant signal reconstruction, moving the VL position away or towards the source caused the perception of source distance to change consistently; however, beyond $\sim 2$m from the source, the change in source distance is not noticeable. For the case of only directional component reconstruction, a decrease in the strength of the signal is noticed with increased source distance, and the sound is heard feebly.

\section{Conclusions}
	The joint estimation using directional filtering and MCLP in the dMCLP scheme is found to be effective for the estimation of direct and diffuse components, along with the source DoA. Ad-hoc distribution of microphone arrays permits inconspicuous placement, and is shown to be effective for source position estimation. The parameterization of the reverberant signal in terms of the directional, diffuse component signals and the estimated source position, is found to be effective for a spatial reconstruction at the receiver through a selection of the virtual listener position. Also, scene modification at the receiver, independent of the transmitter spatial analysis, is possible. Compared to the beamforming based approach, the perceived quality of spatial reconstruction is better even with a smaller number of microphone elements at each array node.

\bibliographystyle{IEEEtran}
\bibliography{references}
\end{document}